\documentclass[journal]{IEEEtran}
\IEEEoverridecommandlockouts
\usepackage{cite}
\usepackage{amsmath,amssymb,amsfonts}
\usepackage{algorithmic}
\usepackage{caption}
\usepackage{graphicx}
\usepackage{textcomp}
\usepackage{multirow}
\usepackage[flushleft]{threeparttable}
\usepackage[table]{xcolor}
\usepackage{xcolor}
\usepackage[hidelinks]{hyperref}
\usepackage{balance}

\usepackage[a4paper, total={184mm,239mm}]{geometry}

\def\BibTeX{{\rm B\kern-.05em{\sc i\kern-.025em b}\kern-.08em
		T\kern-.1667em\lower.7ex\hbox{E}\kern-.125emX}}

\begin{document}



\title{Stress Detection using Context-Aware\\ Sensor Fusion from Wearable Devices}

\author{Nafiul~Rashid,~\IEEEmembership{Student Member,~IEEE,}
    Trier Mortlock,
    and~Mohammad~Abdullah~Al~Faruque,~\IEEEmembership{Senior~Member,~IEEE}
    \thanks{ Nafiul Rashid is a PhD candidate in the Department of Electrical Engineering
        and Computer Science; Trier Mortlock is a PhD candidate in the Department of Mechanical and Aerospace Engineering, and Mohammad Abdullah Al Faruque is a professor in both Electrical Engineering
        and Computer Science and Mechanical and Aerospace Engineering at the University of California, Irvine, CA
        92697, USA, e-mail: (nafiulr@uci.edu)}
    \thanks{ Copyright (c) 2023 IEEE. Personal use of this material is permitted. However, permission to use this material for any other purposes must be obtained from the IEEE by sending a request to pubs-permissions@ieee.org.}
}

\markboth{IEEE Internet of Things Journal,~Vol.~XX, No.~XX, XXXX~2023}%
{Shell \MakeLowercase{\textit{et al.}}: Bare Demo of IEEEtran.cls for IEEE Journals}

\maketitle

\begin{abstract}
Wearable medical technology has become increasingly popular in recent years. One function of wearable health devices is stress detection, which relies on sensor inputs to determine a patient’s mental state. This continuous, real-time monitoring can provide healthcare professionals with vital physiological data and enhance the quality of patient care.
Current methods of stress detection lack: (i) robustness---wearable health sensors contain high levels of measurement noise that degrades performance, and (ii) adaptation---static architectures fail to adapt to changing contexts in sensing conditions.
We propose to address these deficiencies with SELF-CARE, a generalized selective sensor fusion method of stress detection that employs novel techniques of context identification and ensemble machine learning. SELF-CARE uses a learning-based classifier to process sensor features and model the environmental variations in sensing conditions known as the noise context. SELF-CARE uses noise context to selectively fuse different sensor combinations across an ensemble of models to perform robust stress classification. Our findings suggest that for wrist-worn devices, sensors that measure motion are most suitable to understand noise context, while for chest-worn devices, the most suitable sensors are those that detect muscle contraction.
We demonstrate SELF-CARE’s state-of-the-art performance on the WESAD dataset. Using wrist-based sensors, SELF-CARE achieves 86.34\% and 94.12\% accuracy for the 3-class and 2-class stress classification problems, respectively. For chest-based wearable sensors, SELF-CARE achieves 86.19\% (3-class) and 93.68\% (2-class) classification accuracy. This work demonstrates the benefits of utilizing selective, context-aware sensor fusion in mobile health sensing that can be applied broadly to Internet of Things applications.
\end{abstract}

\begin{IEEEkeywords}
Stress Detection, Context-Aware Models, Wearable Health Sensor Fusion, Ensemble Learning.
\end{IEEEkeywords}

\section{Introduction}
\label{sec:intro}




{Advancement in technology and the prevalence of Internet of Things (IoT) has led to the wide adoption of wearable medical devices in recent years.} 
{Wearable medical devices have shaped the study and practice of healthcare by allowing continuous, remote monitoring of vital physiological signs.}
{Wearable health devices can also be used for stress detection, which uses inputs from body-worn sensors to analyze a patient's mental state.}
{Stress detection is of growing interest as recently the American Psychological Association issued a warning about long-term physical and mental health impacts due to stresses from the COVID-19 Pandemic, deeming it a \textit{`a national mental health crisis'} \cite{american2020stress}.}


{Medically, stress is a physiological state that can be triggered by hormonal surges during moments of physical, cognitive, or emotional challenges \cite{goldstein2010adrenal}.}
Stress detection falls under the umbrella of \textit{affective computing}---the area of computing that allows machines to recognize and interpret human emotions \cite{picard2000affective}. 
{Affective computing using wearable devices is a rapidly developing industry, the value of which is projected to expand from \$29 billion to \$140 billion---an increase of nearly five times---by 2025 \cite{affective_computing_market}.}




\subsection{{Research Challenges}}
{The increasing prevalence of wearable health technology---and the data that can be gleaned from this technology---has given rise to a body of academic literature focusing on stress detection \cite{kim2008emotion,gjoreski2016continuous,hovsepian2015cstress,healey2005detecting,picard2001toward,koelstra2011deap,schmidt2018introducing}.}
The relationship between this sensor data and stress states is not governed by known physical equations. {As a result, researchers have used classical machine learning models (e.g., random forests, decision trees) or deep learning models (e.g., convolutional neural networks, long short-term memory) to perform stress classification via supervised learning over labeled datasets with annotated stress states \cite{samyoun2020stress, huynh2021stressnas, rashid2021feature, ragav2019scalable}.}
Deep learning models have benefits in their ability to incorporate temporal modeling from the sensor data into the stress detection problem.
Despite this, in stress detection, classical machine learning models have been more widely adopted compared to deep learning models due to the classical models' lower complexity levels, important for wearable on-device deployment \cite{schmidt2019multi}.  
{However, both of these types of learning-based methods lack robustness when using single sensor modalities, since the coverage area of each sensing modality is limited by the domain in which the sensors operate \cite{naeini2021amser}.}



\begin{figure*}[t]
    \centering
    \includegraphics[width=\linewidth]{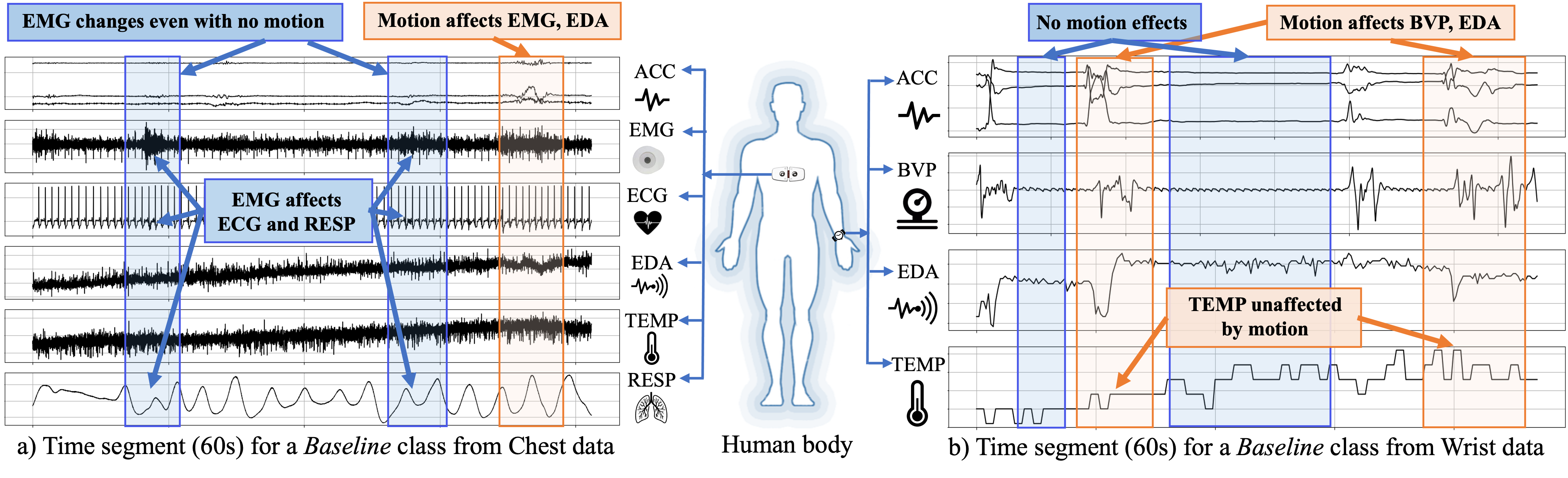}
    \caption{The context of noise from sensor measurements depends on the respective sensor locations on the human body. a) Physiological signals from chest sensors. A baseline segment where EMG affects ECG and RESP even with no motion, whereas ECG remains unaffected even during motion. This shows that EMG is more suitable than ACC to understand the noise context from chest wearable devices. b) Physiological signals from wrist sensors. A baseline segment where BVP and EDA is affected due to motion. Hence, ACC is more suitable  to understand the noise context in wrist wearable devices. Both sets of data in a) and b) are taken from wrist and chest sensors on one subject from the WESAD dataset\cite{schmidt2018introducing}.}
    \label{fig:mot_wrist_chest}
\end{figure*}

Researchers commonly use sensor fusion across multi-modal physiological data to increase the performance of emotion recognition from wearable devices \cite{bota2020emotion}. {\textit{Early} fusion (also known as feature-level fusion) focuses on combining data at the raw-data level. Alternatively, \textit{late} fusion (also known as decision-level fusion) combines the final outputs of a system.} 
Current methods of sensor fusion that employ combinations of early and late fusion still have limited efficacy due to the use of static architectures that cannot adapt to changing sensing conditions within the environment \cite{malawade2022hydrafusion}. 

Another notable challenge in using data from these physiological signals for affective computing is that the data may be susceptible to substantial amounts of sensor noise due to physical motion or muscle contraction.
Throughout the remainder of this paper, we define the noise context of wearable health sensors as the group of external factors that can influence the variation in measurements and noise levels of the sensors. 
This context can be interpreted through intra-sensor relationships in the device as well as through sensing conditions surrounding the device {(\textit{e.g.}, the location of a wearable sensor on the body)}. {And fusing data from multiple sensors without understanding the noise context may lead to performance degradation as found in \cite{schmidt2018introducing}.} 

{The main research challenges we address in this work include: (i) how to effectively fuse multi-modal sensor data from wearable devices; (ii) how to develop an adaptive architecture to account for variations in sensing conditions; and (iii) how to model noise context in wearable sensors to improve stress classification performance.}


\subsection{{Motivation}}
{In this subsection, we provide motivation and qualitative analysis regarding the challenges our approach addresses.}  
Fig. \ref{fig:mot_wrist_chest} shows that the context of noise on sensors varies depending on the location of the wearable device.
Fusing such noisy measurements can subsequently degrade the classification performance \cite{schmidt2019multi}. {For example, Fig. \ref{fig:mot_wrist_chest} b) represents a baseline segment of data from four wrist sensing modalities: tri-axis accelerometer (ACC), blood volume pulse (BVP), electrodermal activity(EDA), skin temperature (TEMP). At several times during the segment, significant motion causes two of the sensors (BVP, EDA) to vary in their readings, which could cause a model to classify this segment \textit{incorrectly} as stress.} {Therefore, it is important to understand the noise context when making sensor fusion decisions. Moreover, it also shows that motion sensors (ACC) have benefits for modeling the noise context in wrist-worn devices.}

On the other hand, Fig. \ref{fig:mot_wrist_chest} a) shows data from six sensing modalities from the chest (ACC, electromyography: EMG, electrocardiogram: ECG, EDA, TEMP, respiration: RESP) for a baseline segment of the subject. While chest motion may affect EMG and EDA, it does not affect ECG. {However, EMG may be affected even without any motion when the subject makes muscle contractions without moving.} This may in turn affect ECG and RESP as shown in Fig. \ref{fig:mot_wrist_chest}. Thus, for chest wearable devices, motion is not the best modality to understand the noise context for sensor fusion decisions. Rather, EMG is more suitable for chest-worn devices which is empirically validated later in Section \ref{sec:results}.

{The aforementioned examples motivate us to develop a context-aware sensor fusion technique that utilizes the noise context of wearable devices to make sensor fusion decisions, which will help us to  maintain performance while avoiding misclassification. Moreover, the developed method should be generalizable to both chest and wrist wearable devices as the noise context varies based on the location of wearable devices.} 
Prior work has shown that stress detection using wrist-based wearable devices can be improved by modeling noise context \cite{rashid2022self}, however, the differences in using chest-based wearable devices have yet to be examined. 

\subsection{{Contributions}}
In this paper, we propose SELF-CARE, a generalized stress detection method that utilizes the noise context of wearable devices to perform sensor fusion.
We show that while motion-based noise context understanding works best for wrist-based wearable devices, muscle contraction works best for chest-based wearable devices. Through experimental evaluation, we demonstrate that EMG is better than ACC in understanding the noise context of chest-based wearable devices.

\noindent
The key contributions of this paper are as follows:

\begin{enumerate}

\item {We introduce a generalized selective sensor fusion method, SELF-CARE, for stress detection from wearable health sensors. SELF-CARE implements a novel context identification method that models noise context based on the location of wearable devices (chest or wrist), and utilizes the noise context to dynamically adjust the sensor fusion performed across an ensemble of machine learning classifiers to improve classification performance.}




\item {We empirically demonstrate that noise context varies based on the location of wearable devices through experimentation across nine different wearable sensors.} Our findings suggest that while motion (ACC) is most suitable to understand the noise context in wrist-worn devices, muscle contraction (EMG) is more suitable to determine noise context in chest-worn devices.

\item {We propose a novel late fusion technique for classification over an ensemble of learners using a Kalman filter that incorporates temporal dynamics.}

\item We perform an extensive performance evaluation of the different combinations of sensors from chest and wrist wearable devices for stress detection. This may serve as the benchmark for the research community to understand, evaluate, and compare the impact of sensor fusion in stress detection.  

\item We validate our methodology on the WESAD dataset, showing that SELF-CARE is suitable for wrist-based and chest-based wearable devices and achieves state-of-the-art performance for the 3-class and 2-class stress detection problems.


\end{enumerate}


\subsection{{Paper Organization}}
{The remainder of this paper is structured as follows. In Section \ref{sec:rw}, we discuss related works in stress and emotion detection and sensor fusion. In Section \ref{sec:prob}, we describe the stress classification problem formulation. In Section \ref{sec:meth}, we introduce the methodology of our context-aware, selective sensor fusion approach. In Section \ref{sec:results} we show the results of our approach on a publicly available stress classification dataset. In Section \ref{sec:disc}, we highlight future directions and limitations, and in Section \ref{sec:con}, we provide concluding remarks.}

\section{Related Works}
\label{sec:rw}
{As this paper presents a context-aware sensor fusion technique for stress detection, we consider the related works from stress detection and sensor fusion. Therefore, we categorize the related works into two parts. In Section \ref{sec:stress_emotion_rw}, we present some related works that consider stress and emotion detection using various sensor modalities. We also discuss the availability of the dataset used in the corresponding works. In Section \ref{sec:sensor_fusion_rw}, we present and compare against the works that mainly focus on sensor fusion techniques for stress detection.}

\subsection{Stress and Emotion Detection}
\label{sec:stress_emotion_rw}
{A number of studies \cite{kim2008emotion,gjoreski2016continuous,hovsepian2015cstress} focus on detecting stress or emotion from physiological signals such as electrocardiograms (ECGs), electromyograms (EMGs), blood volume pulse (BVP), respiration (RESP), electrodermal activity (EDA), and skin temperature (TEMP).} However, these datasets are not publicly available. {Among works with publicly available datasets, authors in \cite{healey2005detecting} detect stress while driving a vehicle, while \cite{picard2001toward} and \cite{koelstra2011deap} perform a more complex analysis on subjects' general emotional states. However, these datasets are limited in that they do not include data on both stress and additional emotions simultaneously.}

{Authors in \cite{schmidt2018introducing} created the WESAD (Wearable Stress and Affect Detection) dataset, which includes data on both stress and amusement states from chest- and wrist-worn devices.} Moreover, the authors compare the classification performance of multiple common machine learning methods using chest-worn sensors, wrist-worn sensors, and their combinations. They conclude that: (i) chest sensors perform better, and wrist sensors become redundant and sometimes even decrease performance, (ii) fusing multiple sensor modalities together can improve results, and (iii) the accelerometer can negatively impact classification performance. The third finding supports our claims that modeling the context as a learned abstraction of motion can be beneficial for wearable devices, and that sometimes fusing all available sensors together reduces performance. {Authors in \cite{samyoun2020stress} use the WESAD dataset to present a translation method using a Generative Adversarial Network (GAN) to generate chest sensor features using the wrist sensors. However, the higher computational complexity of GANs, along with the requirement of chest data during training, limits the application for computing on a wrist-worn device. Authors in \cite{rashid2021feature} propose a hybrid convolutional neural network (CNN) architecture that uses both manually extracted and CNN features for classification, but only uses one sensing modality.} 
 {The authors of \cite{lin2019explainable,fouladgar2021cn} and \cite{huynh2021stressnas} explore the feasibility of deep learning models for stress and emotion detection using the WESAD dataset. However, traditional machine learning models are currently favored over deep learning approaches due to deep learning's increased computational complexity and lack of explainability \cite{schmidt2019multi,bota2020emotion}.} 




\subsection{Sensor Fusion}
\label{sec:sensor_fusion_rw}
Sensor fusion has many benefits when applied to both physiological signals and stress recognition \cite{gravina2017multi,bota2020emotion}. {By combining raw-sensor data or features (early fusion), more information can be extracted from sensor measurements than would otherwise be available.} Likewise, using an ensemble method of multiple learners (late fusion) can increase robustness to sensor/classifier errors. {Performing late fusion on the outputs of multiple classifiers can improve performance, as each classifier can be specialized for its particular set of input data \cite{oviatt2018handbook}.}
Traditional late fusion approaches typically use a voting method over the outputs of the classifiers to make a final decision. Other works have also proposed a learned late fusion method, such as the method discussed in \cite{yan2022emotion}. The authors propose an adaptive fusion method, detailing the benefits of using event-related feature extraction techniques along with an adaptive framework. {However, their approach does not consider the noise context of data for the sensor fusion decision as we do in our approach. Additionally, we also show that the noise context varies based on the location of wearable devices, which has not been addressed in their work. Furthermore, although their late fusion is adaptive, their method is static in that it requires a set number of classifiers. Our model, on the other hand, can dynamically adjust the number and type of classifiers used based on the performance-computation trade-off.}

{Lastly, sensor fusion presents additional benefits when fusing time-series data with temporal correlations, like the data present in physiological sensors.} Kalman filters are tools for estimating unknown quantities by iteratively predicting and updating the estimated state of interest \cite{kalman1960new}, which in our case is the predicted class. {Some works propose using Kalman filters to solve classification problems, \cite{pakrashi2019kalman}, while other works do not consider temporal aspects within their formulation. In this work, we present a novel late fusion method using a Kalman filter to take advantage of the temporal dynamics in the stress classification problem.} 





\section{Problem Formulation}
\label{sec:prob}
As discussed in Section \ref{sec:rw}, fusing multiple heterogeneous physiological signals has benefits for stress detection. The main sources of these physiological signals are generally either chest- or wrist-worn wearable devices. Between the two, wrist-worn wearable devices are more prone to noise induced by random movements of hands, and, as shown in Fig. \ref{fig:mot_wrist_chest}, movements create varying impacts on different physiological signals. Fusing such noisy signals often deteriorates the classification performance \cite{schmidt2018introducing}. {On the other hand, chest-worn wearable devices are less prone to random movements due to their location, but signals may be affected or become noisy for other reasons, such as muscle contraction.} Therefore, it is important to understand the context of the noise which varies based on the location of the wearable  devices. Understanding the noise context can help to dynamically select the less impacted signals to be fused, which will  eventually improve the classification performance.
 The problem formulation for stress detection in a selective approach is provided as follows.

For each input segment of sensor data, the goal of a classifier $\phi$ is to utilize the measurements from available sensors, $\mathbf{X}$, to classify the segment, $\mathbf{Y}$:
\begin{equation}
    \mathbf{Y} = \phi(\mathbf{X}) = [p_1, p_2 , \dots p_c]
    \label{eq:Y}
\end{equation}    
    \begin{equation}
    \mathbf{X} = \{\mathbf{X}_i\}_{i=1 \dots s}
 \end{equation}   
where $s$ is the number of available sensors; ${\mathbf{X}_i}$ represents the measurements from sensor $i$; and ${\mathbf{Y}}$ represents the classifier output which is comprised of the probabilities $p$ of the $c$ classes, (e.g., $c=1$: baseline, $c=2$: stress, $c=3$: amusement).
$\phi$ can be implemented via traditional sensor fusion techniques, a machine learning (ML) or deep learning (DL) model, or an ensemble of ML/DL models. 

Since $\mathbf{X}$ represents data from multiple heterogeneous sensing modalities, sensor fusion can be used to fuse the data to provide a better estimate of $\mathbf{Y}$. In early fusion, the raw sensor inputs are fused before being passed through the classifier as follows:
\begin{equation}
    \mathbf{\hat{Y}}  = \phi(\psi(\mathbf{X}_1, \mathbf{X}_2, \dots , \mathbf{X}_s ))
\end{equation}
where $\psi$ represents the function for fusing the different inputs. In contrast, \textit{late fusion}, involves fusing the outputs of an ensemble of sensor-specific classifiers as follows:
\begin{equation}
    \mathbf{\hat{Y}}_1,\, \mathbf{\hat{Y}}_2,\, \dots, \mathbf{\hat{Y}}_s = \phi_1(\mathbf{X}_1),\,  \phi_2(\mathbf{X}_2),\, \dots , \, \phi_s(\mathbf{X}_s)
\end{equation}
\begin{equation}
    \mathbf{\hat{Y}} = \phi(\mathbf{\hat{Y}}_1, \mathbf{\hat{Y}}_2, \dots , \mathbf{\hat{Y}}_s)
\end{equation}

The context of noise can vary dramatically based  on the wearable device location  and may have a range of impacts on different sensor modalities. This variance calls for the use of an adaptive $\phi$ that selects the sensor modalities to be fused based on the noise context---for example, movements of hands in wrist-worn wearable devices or muscle contractions in chest-worn wearable devices. In this case, $\phi$ represents an ensemble of classification models, and $\phi^*$ represents the selected best subset of models in the ensemble for a given input $\mathbf{X}$.
The context of the noise (either learned and modeled from the inputs or provided externally) is denoted as $\Omega$. We introduce the context identification problem formulation as:
\begin{equation}
    \Omega = \pi(\mathbf{X}) ,
\end{equation}
\begin{equation}
     \phi^* = \rho(\Omega) ,
\end{equation}
where $\pi$ represents a gating model that performs context identification, and $\rho$ represents the mechanism for selecting $\phi^*$ given the identified context $\Omega$.
The goal of $\pi$ and $\rho$ is to select the optimal subset of branch models $\phi^*$ for the inferred context $\Omega$ to maximize stress classification performance for a given $\mathbf{X}$. In our specific case, context is defined as motion for wrist-worn wearable devices or as muscle contraction for chest-worn wearable devices. The inputs to $\pi$ typically consist of measurements from the accelerometer  (wrist-worn) or EMG (chest-worn) based on the wearable device location. 

\section{Methodology}
\label{sec:meth}
\begin{figure}
    \centering
    \includegraphics[width=\linewidth]{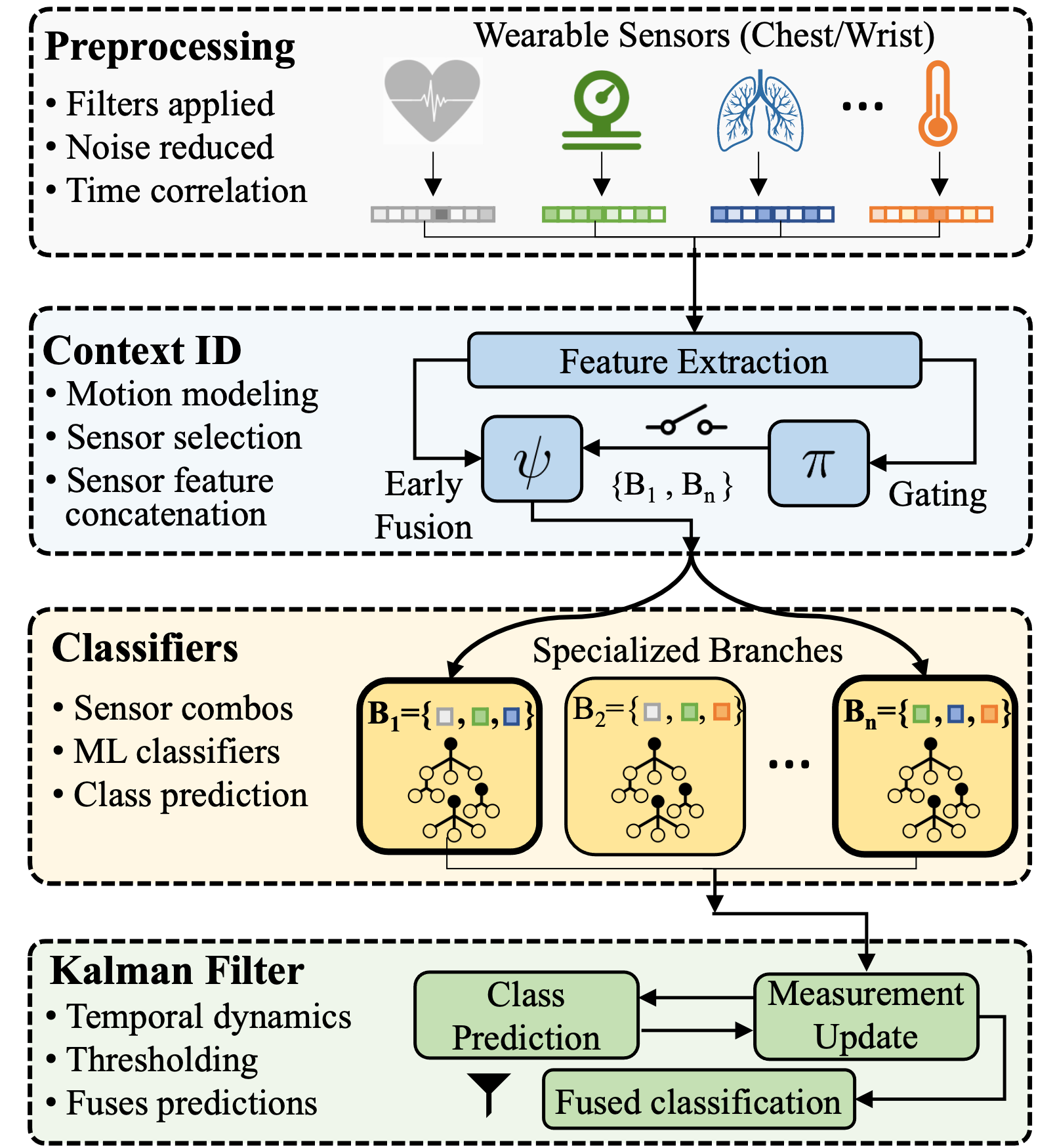}
    \caption{Proposed SELF-CARE Architecture. In this depiction different types of chest/wrist-worn sensors are used, the gating model selects the branches given the context, a Random Forest/AdaBoost classifier is used for the branch models, and a Kalman filter is used for the late fusion over the selected branches.}
    \label{fig:arch}
\end{figure}

In this Section we detail our method, SELF-CARE, depicted in Fig. \ref{fig:arch}. Our method performs stress classification given input sensor measurements from a specified time segment using four main blocks: (i) preprocessing, (ii) context identification, (iii) branch classifiers, and (iv) late fusion. SELF-CARE takes the form of a multi-branched architecture in which different ``branches'' represent stress detection classifiers using different combinations of sensors. {Context identification selects which branches to execute, while late fusion is used to fuse the stress classification predictions if multiple branches are selected.}
The following subsections provide further details on the proposed method.

\subsection{Preprocessing Step}
{SELF-CARE can take in data from varying numbers of heterogeneous or homogeneous physiological sensors as inputs.} Preprocessing is a common step when dealing with raw, unfiltered sensor data. By applying various filters (e.g., band-pass filters or lowpass filters) to the input data, random noises are reduced, and important features are more easily extracted. The preprocessing performed over each sensing modality is detailed in Section \ref{sec:results}. 



\subsection{Context Identification}
\subsubsection{Feature Extraction}
The purpose of the context identification block is to predict which branch classifier(s) will perform the best given an input set of sensor features that are used to model the context of the system. Contextual modeling can help illuminate the performance of various sensors in terms of their levels of noise under different situations and the locations of the wearable device on the human body. 
For wrist wearable devices, we use motion to model the context. Therefore, for wrist devices, we first extract only ACC features as they are directly related to the relative motion of the test subject. {For chest wearable devices, on the other hand, the context is best modeled by muscle contraction, which is captured by EMG signal. We then extract EMG features for chest-worn devices for contextual modeling. Next, these features are processed by the gating model to select the best performing branch.} The feature extraction of the other modalities takes place after the gating model has selected which branch(es) will be used for classification. 
\subsubsection{Gating Model ($\pi$)}
The gating model trains a classifier that uses the ACC/EMG features as inputs to select one of the available branch classifiers according to wrist/chest-worn devices. For wrist-worn device, we shortlist these three branches:$WB_{1}$=\{BVP, EDA, TEMP\}; $WB_{2}$=\{ACC, BVP, EDA\}; $WB_{3}$=\{BVP, EDA\} using Random Forest classifier for both 3-class and 2-class classification. Similarly, for chest-worn devices, we shortlist five branches for 3-class and 2-class classification using AdaBoost classifiers. For 3-class classification, the shortlisted branches are: $CB_{1}$=\{ECG, RESP, EMG, EDA, TEMP\}; $CB_{12}$=\{ECG, EMG, EDA, TEMP\}; $CB_{14}$=\{RESP, EMG, EDA, TEMP\}; $CB_{24}$=\{ECG, EMG, EDA\}; $CB_{27}$=\{EMG, EDA, TEMP\}. For 2-class classification, the shortlisted branches are: $CB_{5}$=\{ACC, ECG, RESP, EDA\}; $CB_{7}$=\{ACC, ECG, EMG, EDA\}; $CB_{9}$=\{ACC, ECG, EDA, TEMP\}; $CB_{13}$=\{ECG, RESP, EDA, TEMP\}; $CB_{20}$=\{ACC, ECG, EDA\}. {The process for choosing these branches is discussed further in Section \ref{SELF-CARE_implementation}. We employ a Decision Tree (DT) classifier for our gating model because it is lightweight and adds minimum overhead to our architecture.}
\subsubsection{Performance-Computation Trade-off ($\delta$)}
\label{sec:delta}
An important feature of SELF-CARE is its ability to balance constraints between performance and computation. We introduce the term $\delta$ that aids the gating decision in considering this trade-off. The gating model outputs prediction probabilities for the available branches with $\bar{b}$ representing the maximum probability branch. $\delta$ has a range between 0 and 1, representing the range in which non-maximum branches are selected by allowing branches with probabilities greater than $\bar{b}-\delta$ to be also selected. Lower $\delta$ values indicate tighter computation constraints, with $\delta=0$ indicating that only the highest probability branch from the gating classifier is selected, while higher $\delta$ values allow more branches to be selected, with $\delta=1$ indicating that all possible branches are selected. 
\subsubsection{Early Fusion ($\psi$)}
Once the branches are selected after applying $\delta$ on the gating model decision, the features for those branches will be extracted and concatenated together to be passed to the corresponding classifiers. For example, while using wrist modalities, if $WB_1$ and $WB_3$ are the selected branches by the gating model for either 3-class  or 2-class classification, the features from BVP, EDA, and TEMP signals are concatenated together using early fusion for $WB_1$, while features from BVP and EDA are fused for branch $WB_{3}$. {Similarly, for 3-class classification using chest modalities, the features from ECG, RESP, EMG, EDA, and TEMP are fused together if the gating model selects the $CB_1$ branch.}

\subsection{Branch Classifiers}
Next, the corresponding branch classifier(s) are used to classify the segment. For our approach, we use a Random Forest (RF) classifier for all three branches of wrist modalities for 3-class and 2-class classification. For chest modalities, we use the AdaBoost classifier for all five branches for 3-class and 2-class classification. The details of the classifier training and selection are provided below in Section \ref{SELF-CARE_implementation}. {Currently, SELF-CARE operates using either only wrist sensors or only chest sensors, however, our method is capable of integrating both sets of branches with modifications to the context identification module.
Each selected branch produces a classification prediction to serve as input for the late fusion method.}


\subsection{Late Fusion Method}
\label{sec:lfb}
The late fusion method is tasked with fusing the class predictions from the various selected branches, $\{\mathbf{\hat{Y}}_1, \mathbf{\hat{Y}}_2, \dots , \mathbf{\hat{Y}}_s\}$, with the goal of producing higher accuracy predictions than any one individual branch by itself. 
{Here we present our \textit{Kalman filter-based} method for classification over an ensemble of classifiers.}
Kalman filters are powerful and commonly used tools for sensor fusion and the broader field of estimation. They are designed to estimate the unknown state of a system along with the state's uncertainty by performing a series of recursive predictions and measurement updates. In the context of our problem, we consider a Kalman filter approach towards the multi-class classification problem like in \cite{pakrashi2019kalman}, and we additionally model the temporal dynamics in the stress classification problem for each sample at time $k$. The general form of the discretized linear dynamics of a system with state $\mathbf{x}$ and measurements $\mathbf{z}$ is given as:
\begin{equation}
\mathbf{x}(k) = \mathbf{F}x(k-1) + \mathbf{v}(k)
\end{equation}
\begin{equation}
\mathbf{z}(k) = \mathbf{H}x(k) + \mathbf{w}(k) 
\end{equation}
where $\mathbf{F}$ is the state transition matrix; $\mathbf{v}$ is the process noise vector, which is modeled as zero-mean, normally distributed random variable with covariance, $\mathbf{Q}$; $\mathbf{H}$ is the measurement matrix relating the state to the measurements; and $\mathbf{w}$ is the measurement noise vector, which also is zero-mean with a normal distribution and covariance, $\mathbf{R}$.

During the prediction step of the Kalman filter, the state estimate and its estimation error covariance matrix, $\mathbf{P}(k)$, are propagated forward through the dynamics model with the added process noise. This step enforces the temporal dependency that the stress class probabilities at the current time step have on the future time step. The prediction equations are:
\begin{equation}
\mathbf{x}(k|k-1) = \mathbf{F}\,\mathbf{x}(k-1|k-1) , 
\end{equation}
\begin{equation}
\mathbf{P}(k|k-1) = \mathbf{F}\,\mathbf{P}(k-1|k-1)\,\mathbf{F}^\top + \mathbf{Q}(k-1) 
\end{equation}
where the notation $(k+1|k)$ indicates the next time step given the current time step. Next, during the update step, measurements are processed and updated estimates of the states and their covariance are corrected according to the measurements. The measurement update equations are as follows:
\begin{equation}
\mathbf{x}(k|k) = \mathbf{x}(k|k-1) -  \mathbf{K}(k)
[\mathbf{H}(\mathbf{x}(k|k-1) - \mathbf{z}(k)] 
\end{equation}
\begin{equation}
\mathbf{P}(k|k) = \mathbf{P}(k|k-1) - \mathbf{K}(k) \mathbf{H} \mathbf{P}(k|k-1)  
\end{equation}
\begin{equation}
\mathbf{K}(k) = \mathbf{P}(k|k-1)  \mathbf{H}^\top  [\mathbf{H}\mathbf{P}(k|k-1) \mathbf{H}^\top + \mathbf{R}(k)]^{-1} 
\end{equation}
with $\mathbf{K}$ representing the Kalman gain. The prediction and update step are iterated to produce an estimate of the state, $\mathbf{x}$, and its associated estimation error covariance, $\mathbf{P}$, representing the uncertainty involved with the state estimate.

For our case, we abstract the multi-class classification problem as follows. The unknown state our filter is attempting to estimate is the probability of each class during each segment. Thus, $\mathbf{x}$ is a $c$ dimensional vector of estimated class probabilities. Additionally, the predictions from each separate classifier are the measurements $\mathbf{z}$, which are processed sequentially per time step. This allows for $s^*$ measurement updates per iteration where $s^*$ is adaptively selected per sample by the gating model. We additionally provide some measurement thresholding during the filter updates that are detailed in Section \ref{sec:kf_tune}. Finally, we arrive at our late fusion output using the Kalman filter-based method:
\begin{equation}
    \mathbf{\hat{Y}}_{kf} = \operatorname*{arg\,max}_c \mathbf{x} 
\end{equation}
where $\mathbf{x}$ is the state vector from the Kalman filter. 
To validate our Kalman-filter based method, we benchmark its performance against commonly used voting mechanisms for late fusion: \textit{hard-voting} and \textit{soft-voting} \cite{oviatt2018handbook}. The method of hard-voting assigns the final class based on the class most commonly voted by each classifier, whereas soft-voting selects the class with the highest average value across all the classifiers. Our results comparing different late fusion approaches are presented in Figures \ref{fig:overall_performance_wrist_3_class}, \ref{fig:overall_performance_wrist_2_class}, \ref{fig:overall_performance_chest_3_class}, and  \ref{fig:overall_performance_chest_2_class} of Section \ref{sec:results}.






\section{Experimental Analysis}
\label{sec:results}
This section presents the experimental findings of SELF-CARE on a wearable health device stress detection dataset. First, we describe the dataset used for evaluation. Second, we explain the training and implementation of our models. Third, we describe our evaluation metrics and analyze experimental results.

\subsection{Dataset}
SELF-CARE is validated on the publicly available WESAD dataset \cite{schmidt2018introducing}. This dataset was selected because it contains data from both wrist- and chest-worn wearable devices, which makes it an ideal dataset for understanding the noise context devices worn on different parts of the body. The dataset contains data for a total of 15 subjects, from both chest- (RespiBAN) and wrist- (Empatica E4) worn sensors. The chest sensors used in RespiBAN  are ACC, ECG, RESP, EMG, EDA, TEMP. The wrist sensors from the Empatica E4 are ACC BVP, EDA, TEMP. 
The dataset has three types of classes related to emotional states: (i) baseline (neutral), (ii) amusement, and (iii) stress. For the 2-class problem, baseline and amusement are grouped together in the non-stress class.


\begin{figure}
    \centering
    \includegraphics[width=\linewidth]{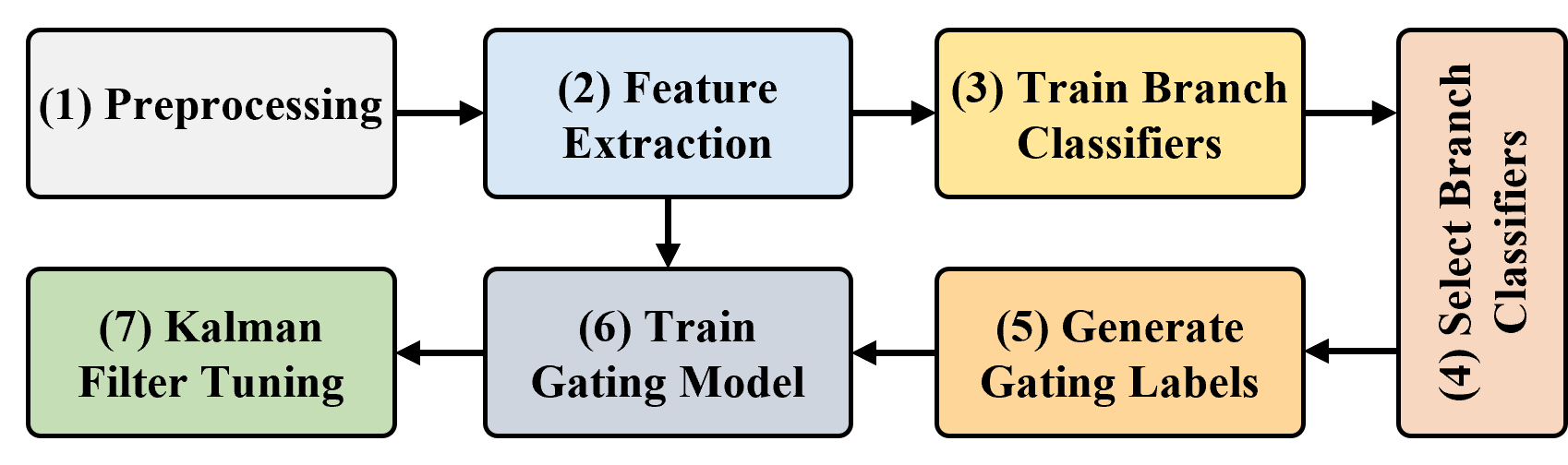}
    \caption{SELF-CARE training and implementation procedure. We follow the steps sequentially as numbered in the figure. To train the gating model, we first generate the gating labels from step 5 and then use extracted features from step 2 to train it.}
    \label{fig:self-care_training}
\end{figure}

\subsection{SELF-CARE Training and Implementation}
\label{SELF-CARE_implementation}
This section describes the training and implementation details for the SELF-CARE architecture, shown in Fig. \ref{fig:self-care_training}. 

\subsubsection{Preprocessing Step} The preprocessing step involves raw data processing to filter out typical noises.

\textbf{Wrist Modalities}: The ACC data is passed through a Finite Impulse Response (FIR) filter with a length of 64 with a cut-off frequency of 0.4 Hz. Following the work in \cite{rashid2021feature}, the raw BVP signal is filtered by a Butterworth band-pass filter of order 3 with cutoff frequencies ($f_1$=0.7 Hz and $f_2$=3.7 Hz), which takes into account the heart rate at rest ($\approx$40 BPM) or high heart rate due to exercise scenarios or tachycardia ($\approx$220 BPM) \cite{salehizadeh2016novel}. The raw EDA signals are filtered using a Butterworth lowpass filter of order 6 with cut-off frequency of 1 Hz. Finally, we use a Savitzky-Golay filter (window size=11, order=3) to smooth the raw TEMP signals.

\textbf{Chest Modalities}: Because the chest data is collected at a very high sampling rate (700 Hz), the signals are first smoothed using  a Savitzky-Golay filter. The ACC data is smoothed using a window size of 31 with an order of 5. The other signals (ECG, EMG, EDA, RESP, and  TEMP) are smoothed using a window size of 11 and an order of 3. Similar to wrist BVP,  the ECG signal is further filtered by a Butterworth band-pass filter of order 3 with cutoff frequencies ($f_1$=0.7 Hz and $f_2$=3.7 Hz) that takes into account the heart rate at rest ($\approx$40 BPM) or high heart rate due to exercise scenarios or tachycardia ($\approx$220 BPM) \cite{salehizadeh2016novel}. The EDA signals are filtered using a Butterworth lowpass filter of order 2 with a cutoff frequency of 5 Hz. To extract some of the peak features (number of peaks, peak amplitude), the EMG signal is passed through a Butterworth lowpass filter of order 3 and a cutoff frequency of 0.5 Hz. We extract other EMG features from the smoothed EMG signal. The RESP signal is filtered by a Butterworth bandpass filter  of order 3 with cutoff frequencies $f_1$=0.1 Hz and $f_2$=0.35 Hz.

The filtered signals from both the wrist and chest  are segmented by a window of 60 seconds of data with a sliding length of 5 seconds following \cite{samyoun2020stress}. This process produces a total of 6458 segments for each signal across all subjects of the WESAD dataset.

  	\begin{table}[t]
   	\centering
   	\begin{threeparttable}
   		\caption{List of Extracted Features}
   		\label{tab:extracted_features}
   		\begin{tabular}{|c|c|}
   				\hline
   				\textbf{Feature Symbol} & \textbf{Feature Names}  \\
   				\hline\hline
       \multicolumn{2}{|c|}{\textbf{ACC Features}}\\
                \hline
   				$\mu_{ACC,i}$, $\sigma_{ACC,i}$ & Mean and STD of each\\
       $i\in\{x, y, z, 3D\}$ & axis and summed over  all axes\\
   				\hline
       $|\int_{ACC,i}|,~i\in\{x, y, z, 3D\}$ & Absolute integral for each/all axes\\
   				\hline
   				$f^{peak}_{ACC,j},~j\in\{x, y, z\}$ & Peak  frequency of each axis\\
   			\hline\hline
                \multicolumn{2}{|c|}{\textbf{ECG/BVP Features}}\\
                \hline
   				$\mu_{HR}$, $\sigma_{HR}$ & Mean and STD of HR \\
   				\hline
   				$\mu_{HRV}$, $\sigma_{HRV}$ & Mean and STD of HRV \\
   				\hline
   				\multirow{2}{*}{$NN50$, $pNN50$} & Number and percentage of HRV \\
   				&intervals differing more than 50 ms  \\
   				\hline
   				$rms_{HRV}$ & Root mean square of the HRV\\
   				\hline
   				$f^{x}_{HRV}$& Energy in ultra-low,  low, high,\\
   				$x\in{ULF, LF, HF, UHF}$& ultra-high frequency band of the HRV\\
   				\hline
   				$f^{LF/HF}_{HRV}$ & Ratio of LF and HF component\\
   				\hline
   				$\sum_{x}^{f}$ & $\sum$ of the frequency components\\
   				$x\in{ULF, LF, HF, UHF}$& in ULF-HF\\
   				\hline
   				$rel_{x}^{f}$ & Relative power of freq. components\\
   				\hline
   				$LF_{norm}$, $HF_{norm}$ & Normalised LF and HF
component \\
\hline\hline
       \multicolumn{2}{|c|}{\textbf{EMG Features}}\\
                \hline
   				$\mu_{EDA}$, $\sigma_{EDA}$ & Mean and STD of EMG\\
       
   				 \hline
       $range_{EDA}$ & Dynamic range  of EMG\\
       \hline
       $|\int_{EMG}|$ & Absolute integral\\
       \hline
       $\Bar{\pi}_{EMG}$ & Median of EMG\\
       \hline
       $P_{EMG}^{10},~P_{EMG}^{90}$ & $10^{th}$ and $90^{th}$ percentile\\
       \hline
       $\mu_{EMG}^{f}$, $\Bar{f}_{EMG}$, $f_{EMG}^{peak}$  & Mean, median, and peak frequency\\
       \hline
       PSD($f_{EMG}$)  & Energy in seven bands\\
       \hline
       $\#^{peaks}_{EMG}$ & Number of peaks\\
       \hline
       $\mu_{EMG}^{amp},~\sigma_{EMG}^{amp}$  & Mean and STD of peak amplitude \\
       \hline
       $\sum_{EMG}^{amp},~\Bar{\sum}_{EMG}^{amp}$  & $\sum$ and norm. $\sum$ of peak amplitude \\
   				\hline\hline
       \multicolumn{2}{|c|}{\textbf{EDA Features}}\\
                \hline
   				$\mu_{EDA}$, $\sigma_{EDA}$ & Mean and STD of EDA\\
       \hline
       $min_{EDA},~max_{EDA}$ & Min and max  value of EDA\\
   				 \hline
       $\delta_{EDA},~range_{EDA}$ & Slope and dynamic range  of EDA\\
       \hline
       $\mu_{SCL}$, $\sigma_{SCL}$, $\sigma_{SCR}$ & Mean and STD of SCL/SCR\\
       \hline
       $Corr_{SCL,t}$ & Correlation between SCL and time\\
       \hline
       $\#~SCR$ & Number of SCR segments\\
       \hline
       $\sum_{SCR}^{amp},~\sum_{SCR}^{t}$  & $\sum$ of SCR magnitudes and duration \\
       \hline
       $\int_{SCR}$  & Area under SCR segments \\
       \hline\hline
       \multicolumn{2}{|c|}{\textbf{RESP Features}}\\
                \hline
   				$\mu_{x}$, $\sigma_{x}$ & Mean and STD of inhalation (I)\\
       $x\in{I,E}$ & exhalation (E) duration\\
       \hline
       I/E & Inhalation/exhalation ratio\\
       \hline
       $vol_{insp},~range_{RESP}$ & Volume and range  of RESP\\
       \hline
       $rate_{RESP},~ \sum_{RESP}$ & Respiration rate and duration\\
       \hline\hline
       \multicolumn{2}{|c|}{\textbf{TEMP Features}}\\
                \hline
   				$\mu_{TEMP}$, $\sigma_{TEMP}$ & Mean and STD of TEMP\\
       \hline
    
       $min_{TEMP},~max_{TEMP}$ & Min and max  of TEMP\\
       \hline
       $\delta_{TEMP},~range_{TEMP}$ & Slope and dynamic range  of TEMP\\
       \hline
    
   			\end{tabular}
  			\begin{tablenotes}
			\item \textit{Standard Deviation (STD), Skin Conductance Response (SCR), Skin Conductance Level  (SCL), Heart Rate (HR), Heart Rate Variability (HRV)}
			\end{tablenotes}
   		\end{threeparttable}
   	\end{table}

\subsubsection{Feature Extraction}
We extract the same wrist and chest sensor features as used in \cite{schmidt2018introducing}, some of which include mean/standard deviations, correlations, slope, and dynamic ranges, peak and power frequencies, and absolute integrals.
We note that this feature extraction is only performed across the sensors that are selected to run by the gate for a given input sample.  Table \ref{tab:extracted_features}  contains the list  of extracted features. We refer readers to \cite{schmidt2018introducing} for further details of extracted features per sensor. 

\subsubsection{Train Branch Classifiers}
To train the individual branch classifiers within SELF-CARE, we train using different combinations of input sensor data.

For Wrist Modalities, we use five different early fusion combinations of wrist sensors as input branches: $WB_{1}$=\{BVP, EDA, TEMP\}; $WB_{2}$=\{ACC, BVP, EDA\}; $WB_{3}$=\{BVP, EDA\}; $WB_{4}$=\{ACC, BVP\}; $WB_{5}$=\{ACC, EDA\} as shown in  Tables \ref{tab:wrist_3_class_analysis} and \ref{tab:wrist_2_class_analysis}. For chest modalities, we tried forty-two different combinations of chest sensors as input branches as shown in Tables \ref{tab:chest_3_class_analysis} and \ref{tab:chest_2_class_analysis}.

We evaluate each branch on five different machine learning classifiers---Decision Tree (DT), Random Forest (RF), AdaBoost (AB), Linear Discriminant Analysis (LDA), K-Nearest Neighbor (KNN). We selected these classifiers to ensure a fair comparison with the original WESAD work \cite{schmidt2018introducing}. 
Following the work in \cite{schmidt2018introducing}, we use the same configurations for the classifiers. We use DT as the base estimator for the RF and AB ensemble classifiers, and use 100 base estimators for both RF and AB. In order to measure the splitting quality of the decision nodes, we used information gain and set the minimum number of samples to split a node to 20. For KNN, the K value is set to 9. All classifiers are trained using leave-one-subject-out (LOSO) validation.

\subsubsection{Select Branch Classifiers}
\label{sec:branch_classifier_selection}
We select the branches with the least amount of training loss to be used. The training loss is calculated from the classification confidence of the trained classifiers on the training samples using the categorical cross-entropy, $CE = - \sum_{i}^{n_c} y_{i} \log \hat{y}_{i}$, where $y$ is the one hot encoded true label of a sample, $\hat{y}$ is the corresponding classification output for that sample, and ${n_c}$ is the number of classes. $CE$ is then calculated for all the training samples across all rounds of LOSO validation.

Next, out of the 25 (5 branches x 5 classifiers per branch) possible branch classifiers for wrist modalities, RF classifiers for input branches $WB_{1}$, $WB_{2}$, and $WB_{3}$ are selected as the branch classifiers for both 3-class and 2-class classification. Similarly, for chest modalities, out of 210 (42 branches x 5 classifiers per branch) possible branches, AB classifiers for input branches $CB_{1}$, $CB_{12}$, $CB_{14}$, $CB_{24}$, and $CB_{27}$ are selected for 3-class classification. And for 2-class classification, we select AB classifiers for input branches $CB_{5}$, $CB_{7}$, $CB_{9}$, $CB_{13}$, and $CB_{20}$ for use within our SELF-CARE methodology. These classifier selections are informed by the extensive experiments we performed across the classifiers variations, which we benchmark in Tables \ref{tab:wrist_3_class_analysis}, \ref{tab:wrist_2_class_analysis},  \ref{tab:chest_3_class_analysis}, and \ref{tab:chest_2_class_analysis}.

\subsubsection{Generate Gating Labels}
\label{sec:generate_gating_label} 
The objective of the gating model is to predict one or a subset of branch classifiers from the classifiers listed in Section \ref{sec:branch_classifier_selection} to be used in our  SELF-CARE methodology.
For each of the training samples, we generate gating labels representing the branch that has the least amount of training loss. These gating labels will be used to train the gating model. For each round of LOSO validation, gating labels are generated based only on the training data, and no test data is used to ensure the validity of our approach.

\begin{table*}[!ht]
	\centering
	\caption{Early Fusion Performance of Wrist Modalities in WESAD Dataset for 3-Class (Baseline vs. Stress vs. Amusement)}
	\label{tab:wrist_3_class_analysis}
	\begin{tabular}{|c||c|c||c|c||c|c||c|c||c|c|}
		\hline
		\multirow{2}{*}{\textbf{Modality Used}} &
		\multicolumn{2}{c||}{\textbf{DT}} &\multicolumn{2}{c||}{\cellcolor{gray!20}\textbf{RF}} & \multicolumn{2}{c||}{\textbf{AB}}  & \multicolumn{2}{c||}{\textbf{LDA}}  & \multicolumn{2}{c|}{\textbf{KNN}} \\
		\cline{2-3}\cline{4-5}\cline{6-7}\cline{8-9}\cline{10-11}
		& \textbf{Mac. F1} & \textbf{Acc.} & \textbf{Mac. F1} & \textbf{Acc.} & \textbf{Mac. F1} & \textbf{Acc.}& \textbf{Mac. F1} & \textbf{Acc.} & \textbf{Mac. F1} & \textbf{Acc.}\\
		\hline\hline

        \cellcolor{gray!20} {\textbf{$WB_{1}$=\{BVP, EDA, TEMP\}}} & 56.23 & 62.32 & 	\cellcolor{gray!20}\textbf{62.73} & \cellcolor{gray!20}\textbf{76.62} & 63.78 & 75.78	  &	52.62 &	61.79 &	58.3 & 69.04\\

		\hline
		\cellcolor{gray!20}{\textbf{$WB_{2}$=\{ACC, BVP, EDA\}}} & 58.46 &	48.27 &	\cellcolor{gray!20}\textbf{62.88} &	\cellcolor{gray!20}\textbf{77.71} &	62.39 &	76.63 &	60.23 &	69.63 &	58.9 &	68.55\\
		\hline
		\cellcolor{gray!20}{\textbf{$WB_{3}$=\{BVP, EDA\}}} & 55.14 &	59.02 &	\cellcolor{gray!20}\textbf{61.02} &	\cellcolor{gray!20}\textbf{73.96} &	60.67 &	72.54 &	56.55 &	69.8 &	65.73 &	53.44\\
		\hline
		\textbf{$WB_{4}$=\{ACC, BVP\}} & 51.54 & 60.66 & 56.86 & 71.38 & 57.83 & 71.96 &	58.67 &	68.36 &	55.51 &	67.05\\
		\hline
		\textbf{$WB_{5}$=\{ACC, EDA\}} & 47.98 & 54.5 & 52.97 & 70.15 & 56.47 &	71.31 & 57.71 &	68.6 &	58.75 &	64.87\\
		\hline
\end{tabular}
\end{table*}

\begin{table*}[t]
	\centering
	\caption{Early Fusion Performance of Wrist Modalities in WESAD Dataset for 2-Class (Stress vs. Non-stress)}
	\label{tab:wrist_2_class_analysis}
	\begin{tabular}{|c||c|c||c|c||c|c||c|c||c|c|}
		\hline
		\multirow{2}{*}{\textbf{Modality Used}} &
		\multicolumn{2}{c||}{\textbf{DT}} &\multicolumn{2}{c||}{\cellcolor{gray!20}\textbf{RF}} & \multicolumn{2}{c||}{\textbf{AB}}  & \multicolumn{2}{c||}{\textbf{LDA}}  & \multicolumn{2}{c|}{\textbf{KNN}} \\
		\cline{2-3}\cline{4-5}\cline{6-7}\cline{8-9}\cline{10-11}
		& \textbf{Mac. F1} & \textbf{Acc.} & \textbf{Mac. F1} & \textbf{Acc.} & \textbf{Mac. F1} & \textbf{Acc.}& \textbf{Mac. F1} & \textbf{Acc.} & \textbf{Mac. F1} & \textbf{Acc.}\\
		\hline\hline
		
		\cellcolor{gray!20} {\textbf{$WB_{1}$=\{BVP, EDA, TEMP\}}} & 74.1	& 84.27 &	\cellcolor{gray!20}\textbf{84.66} &	\cellcolor{gray!20}\textbf{89.01} &	85.29 &	88.96 &	71.46 &	77.32 &	83.74 &	86.56\\
		\hline
		\cellcolor{gray!20}{\textbf{$WB_{2}$=\{ACC, BVP, EDA\}}} & 69.44 & 77.06 & \cellcolor{gray!20}\textbf{85.08} & \cellcolor{gray!20}\textbf{88.76} & 85.44 & 88.45	& 85.66	& 87.92	& 80.25	& 83.62\\
		\hline
		\cellcolor{gray!20}{\textbf{$WB_{3}$=\{BVP, EDA\}}} & 80.8 & 84.48 & \cellcolor{gray!20}\textbf{86.37} & \cellcolor{gray!20}\textbf{89.33} & 86.13 & 89.26 & 83.77 & 86.55 & 79.7 & 83.66\\
		\hline
		\textbf{$WB_{4}$=\{ACC, BVP\}} & 74.97 & 79.94 & 76.43 & 82.45 & 79.77 & 84.21 & 82.37 & 85.07 & 76.49 & 80.13\\
		\hline
		\textbf{$WB_{5}$=\{ACC, EDA\}} & 65.65 & 76.1 & 72.77 & 82.42 & 75.39 & 83.52 & 78.66 & 84.19 & 73.72 & 77.55\\
		\hline
\end{tabular}
\end{table*}

\begin{table*}[!ht]
	\centering
	\caption{Early Fusion Performance of Chest Modalities in WESAD Dataset for 3-Class (Baseline vs. Stress vs. Amusement)}
	\label{tab:chest_3_class_analysis}
	\begin{tabular}{|c||c|c||c|c||c|c||c|c||c|c|}
		\hline
		\multirow{2}{*}{\textbf{Modality Used}} &
		\multicolumn{2}{c||}{\textbf{DT}} &\multicolumn{2}{c||}{\textbf{RF}} & \multicolumn{2}{c||}{\cellcolor{gray!20}\textbf{AB}}  & \multicolumn{2}{c||}{\textbf{LDA}}  & \multicolumn{2}{c|}{\textbf{KNN}} \\
		\cline{2-3}\cline{4-5}\cline{6-7}\cline{8-9}\cline{10-11}
		& \textbf{M. F1} & \textbf{Acc.} & \textbf{M. F1} & \textbf{Acc.} & \textbf{M. F1} & \textbf{Acc.}& \textbf{M. F1} & \textbf{Acc.} & \textbf{M. F1} & \textbf{Acc.}\\
		\hline\hline
\cellcolor{gray!20}{\textbf{$CB_{1}$=\{ECG,RESP,EMG,EDA,TEMP\}}} & 54.26 & 62.07 &	58.68 & 71.39 & \cellcolor{gray!20}\textbf{65.63} & \cellcolor{gray!20}\textbf{76.53} & 30.14 & 38.2 & 53.87 &	65.1\\
\hline
\textbf{$CB_{2}$=\{ACC,ECG,RESP,EMG,TEMP\}} & 49.41 & 57.01 &	53.55 & 69.1 & 60.18 & 71.88 & 55.71 & 72.0 & 43.15 &	51.5\\
\hline\textbf{$CB_{3}$=\{ACC,RESP,EMG,TEMP\}} & 45.8 & 55.14 &	53.46 & 68.58 & 57.11 & 69.14 & 54.93 & 66.07 & 44.79 &	55.49\\
\hline\textbf{$CB_{4}$=\{ACC,ECG,RESP,EMG\}} & 48.29 & 55.55 &	50.37 & 63.36 & 54.52 & 65.55 & 51.8 & 62.8 & 43.04 &	51.56\\
\hline\textbf{$CB_{5}$=\{ACC,ECG,RESP,EDA\}} & 44.33 & 53.75 &	52.32 & 69.15 & 55.02 & 73.09 & 44.05 & 54.93 & 43.73 &	52.43\\
\hline\textbf{$CB_{6}$=\{ACC,ECG,EMG,TEMP\}} & 48.96 & 56.55 &	53.98 & 69.58 & 58.98 & 70.88 & 55.31 & 71.41 & 42.33 &	49.93\\
\hline\textbf{$CB_{7}$=\{ACC,ECG,EMG,EDA\}} & 51.22 & 61.09 &	57.0 & 72.64 & 59.54 & 73.35 & 45.79 & 56.21 & 46.82 &	56.17\\
\hline\textbf{$CB_{8}$=\{ACC,RESP,EDA,TEMP\}} & 42.79 & 51.74 &	54.6 & 71.06 & 52.34 & 67.32 & 22.66 & 24.62 & 40.91 &	48.78\\
\hline\textbf{$CB_{9}$=\{ACC,ECG,EDA,TEMP\}} & 41.94 & 51.1 &	54.75 & 72.57 & 57.36 & 75.99 & 38.59 & 48.37 & 45.53 &	54.2\\
\hline
\textbf{$CB_{10}$=\{ECG,RESP,EMG,TEMP\}} & 47.17 & 52.8 &	58.11 & 70.71 & 61.8 & 71.92 & 54.51 & 68.6 & 51.17 &	59.94\\
\hline
{\textbf{$CB_{11}$=\{ECG,RESP,EMG,EDA\}}} & 51.14 & 59.24 &	57.7 & 68.9 & 60.47 & 70.68 & 50.34 & 60.93 & 51.54 &	62.99\\
\hline
\cellcolor{gray!20}{\textbf{$CB_{12}$=\{ECG,EMG,EDA,TEMP\}}} & 53.95 & 61.41 &	57.95 & 71.12 & \cellcolor{gray!20}\textbf{62.09} & \cellcolor{gray!20}\textbf{74.51} & 31.52 & 39.52 & 53.31 &	63.83\\
\hline
\textbf{$CB_{13}$=\{ECG,RESP,EDA,TEMP\}} & 51.75 & 60.09 &	55.02 & 72.85 & 57.73 & 73.45 & 31.98 & 38.97 & 57.14 &	70.26\\
\hline
\cellcolor{gray!20}{\textbf{$CB_{14}$=\{RESP,EMG,EDA,TEMP\}}} & 48.93 & 54.44 &	60.87 & 71.41 & \cellcolor{gray!20}\textbf{63.68} & \cellcolor{gray!20}\textbf{74.16} & 31.59 & 37.16 & 51.69 &	64.38\\
\hline\textbf{$CB_{15}$=\{ACC,EDA,TEMP\}} & 41.23 & 49.34 &	53.69 & 69.18 & 52.91 & 68.95 & 24.18 & 25.84 & 42.29 &	50.57\\
\hline\textbf{$CB_{16}$=\{ACC,EMG,EDA\}} & 48.81 & 58.07 &	54.59 & 69.32 & 54.7 & 69.25 & 35.84 & 45.83 & 44.99 &	54.87\\
\hline\textbf{$CB_{17}$=\{ACC,RESP,EDA\}} & 43.23 & 51.69 &	51.5 & 67.36 & 49.91 & 66.85 & 35.15 & 45.74 & 39.96 &	49.74\\
\hline\textbf{$CB_{18}$=\{ACC,ECG,RESP\}} & 40.4 & 50.19 &	48.55 & 61.61 & 50.11 & 65.04 & 51.39 & 61.91 & 39.85 &	48.3\\
\hline\textbf{$CB_{19}$=\{ACC,RESP,EMG\}} & 45.19 & 52.93 &	48.03 & 61.65 & 50.84 & 62.98 & 43.41 & 54.94 & 42.54 &	54.79\\
\hline\textbf{$CB_{20}$=\{ACC,ECG,EDA\}} & 44.66 & 53.48 &	53.36 & 69.41 & 53.78 & 72.28 & 43.15 & 53.74 & 42.98 &	50.43\\
\hline\textbf{$CB_{21}$=\{ECG,RESP,EMG\}} & 43.25 & 49.07 &	51.13 & 58.51 & 52.17 & 59.63 & 54.68 & 65.37 & 47.82 &	57.02\\
\hline\textbf{$CB_{22}$=\{ECG,EDA,TEMP\}} & 52.6 & 62.66 &	55.06 & 72.59 & 57.63 & 72.9 & 33.23 & 39.41 & 56.58 &	68.1\\
\hline\textbf{$CB_{23}$=\{ECG,RESP,EDA\}} & 49.02 & 55.73 &	51.74 & 65.97 & 50.34 & 65.1 & 46.24 & 59.24 & 52.38 &	64.95\\
\hline
\cellcolor{gray!20}{\textbf{$CB_{24}$=\{ECG,EMG,EDA\}}} & 52.58 & 60.19 &	57.89 & 68.58 & \cellcolor{gray!20}\textbf{61.69} & \cellcolor{gray!20}\textbf{71.25} & 49.41 & 59.77 & 50.23 &	60.75\\
\hline\textbf{$CB_{25}$=\{RESP,EMG,EDA\}} & 42.09 & 49.83 &	56.13 & 64.04 & 61.46 & 69.09 & 39.07 & 49.63 & 48.25 &	61.33\\
\hline\textbf{$CB_{26}$=\{RESP,EDA,TEMP\}} & 45.95 & 54.85 &	56.76 & 74.1 & 54.56 & 71.05 & 22.48 & 23.51 & 50.27 &	64.98\\
\hline
\cellcolor{gray!20}{\textbf{$CB_{27}$=\{EMG,EDA,TEMP\}}} & 49.94 & 55.45 &	61.32 & 71.51 & \cellcolor{gray!20}\textbf{64.72} & \cellcolor{gray!20}\textbf{74.36} & 32.73 & 40.29 & 51.23 &	62.91\\
\hline
\textbf{$CB_{28}$=\{ACC,RESP\}} & 42.27 & 51.12 &	48.39 & 60.41 & 45.06 & 58.18 & 43.3 & 56.92 & 40.93 &	52.46\\
\hline\textbf{$CB_{29}$=\{ACC,EMG\}} & 44.36 & 52.12 &	47.96 & 61.39 & 50.41 & 63.09 & 42.04 & 53.55 & 42.18 &	53.4\\
\hline\textbf{$CB_{30}$=\{ACC,ECG\}} & 39.41 & 48.76 &	49.48 & 63.37 & 48.79 & 62.9 & 51.14 & 61.68 & 37.45 &	45.01\\
\hline\textbf{$CB_{31}$=\{ACC,EDA\}} & 42.26 & 49.04 &	51.56 & 67.42 & 47.84 & 65.89 & 32.8 & 42.47 & 40.46 &	50.72\\
\hline\textbf{$CB_{32}$=\{ACC,TEMP\}} & 39.86 & 48.43 &	49.58 & 62.29 & 46.81 & 59.66 & 52.05 & 63.16 & 42.78 &	52.02\\
\hline\textbf{$CB_{33}$=\{ECG,RESP\}} & 42.24 & 47.3 &	44.13 & 54.5 & 45.72 & 55.92 & 52.31 & 66.59 & 45.34 &	56.68\\
\hline\textbf{$CB_{34}$=\{ECG,EMG\}} & 40.99 & 46.91 &	51.12 & 57.73 & 50.89 & 58.93 & 54.25 & 64.96 & 48.46 &	55.93\\
\hline\textbf{$CB_{35}$=\{ECG,EDA\}} & 51.3 & 57.86 &	50.4 & 64.85 & 50.02 & 64.46 & 44.76 & 57.57 & 50.21 &	61.31\\
\hline\textbf{$CB_{36}$=\{ECG,TEMP\}} & 41.34 & 48.0 &	48.82 & 63.43 & 51.94 & 63.68 & 53.47 & 69.58 & 50.07 &	59.97\\
\hline\textbf{$CB_{37}$=\{EDA,TEMP\}} & 47.52 & 56.72 &	55.39 & 72.88 & 53.17 & 69.89 & 23.22 & 24.12 & 47.97 &	59.91\\
\hline\textbf{$CB_{38}$=\{RESP,EDA\}} & 39.88 & 47.63 &	50.19 & 60.34 & 45.98 & 54.46 & 30.8 & 42.41 & 48.9 &	64.12\\
\hline\textbf{$CB_{39}$=\{RESP,EMG\}} & 42.73 & 50.93 &	47.94 & 57.68 & 47.65 & 57.34 & 44.9 & 57.34 & 45.69 &	56.73\\
\hline\textbf{$CB_{40}$=\{RESP,TEMP\}} & 41.78 & 53.36 &	51.85 & 69.44 & 49.91 & 59.96 & 56.65 & 71.94 & 48.43 &	62.27\\
\hline\textbf{$CB_{41}$=\{EMG,EDA\}} & 42.9 & 51.26 &	55.5 & 63.64 & 60.54 & 68.68 & 36.4 & 45.9 & 49.92 &	60.11\\
\hline\textbf{$CB_{42}$=\{EMG,TEMP\}} & 40.63 & 45.37 &	62.12 & 71.95 & 63.16 & 70.91 & 58.63 & 67.99 & 50.41 &	59.67\\
\hline
\end{tabular}
\end{table*}

\begin{table*}[t]
	\centering
	\caption{Early Fusion Performance of Chest Modalities in WESAD Dataset for 2-Class (Stress vs. Non-stress)}
	\label{tab:chest_2_class_analysis}
	\begin{tabular}{|c||c|c||c|c||c|c||c|c||c|c|}
		\hline
		\multirow{2}{*}{\textbf{Modality Used}} &
		\multicolumn{2}{c||}{\textbf{DT}} &\multicolumn{2}{c||}{\textbf{RF}} & \multicolumn{2}{c||}{\cellcolor{gray!20}\textbf{AB}}  & \multicolumn{2}{c||}{\textbf{LDA}}  & \multicolumn{2}{c|}{\textbf{KNN}} \\
		\cline{2-3}\cline{4-5}\cline{6-7}\cline{8-9}\cline{10-11}
		& \textbf{M. F1} & \textbf{Acc.} & \textbf{M. F1} & \textbf{Acc.} & \textbf{M. F1} & \textbf{Acc.}& \textbf{M. F1} & \textbf{Acc.} & \textbf{M. F1} & \textbf{Acc.}\\
		\hline\hline
        \textbf{$CB_{1}$=\{ECG,RESP,EMG,EDA,TEMP\}} & 73.17 & 75.85 &	82.02 & 83.24 & 81.45 & 84.14 & 46.32 & 48.77 & 74.68 &	78.81\\
\hline\textbf{$CB_{2}$=\{ACC,ECG,RESP,EMG,TEMP\}} & 67.25 & 73.11 &	80.12 & 83.86 & 77.07 & 82.75 & 77.53 & 79.98 & 63.81 &	69.48\\
\hline\textbf{$CB_{3}$=\{ACC,RESP,EMG,TEMP\}} & 68.85 & 76.61 &	78.15 & 83.44 & 73.16 & 81.56 & 74.03 & 77.17 & 61.97 &	70.88\\
\hline\textbf{$CB_{4}$=\{ACC,ECG,RESP,EMG\}} & 66.93 & 72.16 &	70.57 & 78.06 & 72.98 & 80.57 & 75.66 & 79.83 & 64.12 &	69.86\\
\hline
\cellcolor{gray!20}\textbf{$CB_{5}$=\{ACC,ECG,RESP,EDA\}} & 70.39 & 75.81 &	82.41 & 84.21 & \cellcolor{gray!20}\textbf{83.21} & \cellcolor{gray!20}\textbf{85.64} & 68.46 & 75.02 & 74.6 &	77.75\\
\hline\textbf{$CB_{6}$=\{ACC,ECG,EMG,TEMP\}} & 66.06 & 72.22 &	80.13 & 83.88 & 75.77 & 82.36 & 77.13 & 79.64 & 62.97 &	68.0\\
\hline
\cellcolor{gray!20}\textbf{$CB_{7}$=\{ACC,ECG,EMG,EDA\}} & 72.45 & 77.72 &	83.49 & 85.64 & \cellcolor{gray!20}\textbf{82.29} & \cellcolor{gray!20}\textbf{85.72} & 69.6 & 75.2 & 72.05 &	75.86\\
\hline\textbf{$CB_{8}$=\{ACC,RESP,EDA,TEMP\}} & 68.71 & 73.26 &	81.03 & 84.16 & 72.19 & 79.34 & 29.71 & 32.23 & 66.42 &	74.28\\
\hline
\cellcolor{gray!20}\textbf{$CB_{9}$=\{ACC,ECG,EDA,TEMP\}} & 65.13 & 70.79 &	84.47 & 86.12 & \cellcolor{gray!20}\textbf{82.15} & \cellcolor{gray!20}\textbf{85.2} & 59.5 & 61.72 & 75.02 &	78.4\\
\hline\textbf{$CB_{10}$=\{ECG,RESP,EMG,TEMP\}} & 52.55 & 54.77 &	76.87 & 80.08 & 76.02 & 80.39 & 76.89 & 79.17 & 68.32 &	73.94\\
\hline\textbf{$CB_{11}$=\{ECG,RESP,EMG,EDA\}} & 74.89 & 77.51 &	81.23 & 82.93 & 82.25 & 84.79 & 70.15 & 75.69 & 73.76 &	77.89\\
\hline\textbf{$CB_{12}$=\{ECG,EMG,EDA,TEMP\}} & 72.35 & 75.25 &	82.05 & 83.39 & 79.64 & 82.7 & 47.71 & 50.14 & 73.03 &	77.14\\
\hline
\cellcolor{gray!20}\textbf{$CB_{13}$=\{ECG,RESP,EDA,TEMP\}} & 73.62 & 75.58 &	79.2 & 80.24 & \cellcolor{gray!20}\textbf{83.31} & \cellcolor{gray!20}\textbf{84.78} & 49.66 & 51.9 & 79.22 &	82.04\\
\hline\textbf{$CB_{14}$=\{RESP,EMG,EDA,TEMP\}} & 70.82 & 73.11 &	81.8 & 84.15 & 77.41 & 81.44 & 47.13 & 50.71 & 67.67 &	76.19\\
\hline\textbf{$CB_{15}$=\{ACC,EDA,TEMP\}} & 66.82 & 72.0 &	80.51 & 83.08 & 71.5 & 79.53 & 32.61 & 35.96 & 66.95 &	73.5\\
\hline\textbf{$CB_{16}$=\{ACC,EMG,EDA\}} & 69.7 & 76.11 &	78.93 & 83.67 & 74.92 & 82.07 & 54.66 & 61.85 & 64.89 &	71.65\\
\hline\textbf{$CB_{17}$=\{ACC,RESP,EDA\}} & 66.46 & 73.42 &	77.17 & 81.81 & 70.91 & 79.16 & 53.28 & 62.59 & 63.82 &	71.44\\
\hline\textbf{$CB_{18}$=\{ACC,ECG,RESP\}} & 62.03 & 69.17 &	72.22 & 78.51 & 75.78 & 81.16 & 74.68 & 79.53 & 65.87 &	69.49\\
\hline\textbf{$CB_{19}$=\{ACC,RESP,EMG\}} & 63.76 & 71.48 &	66.44 & 77.19 & 64.21 & 75.5 & 63.29 & 71.03 & 60.84 &	70.04\\
\hline
\cellcolor{gray!20}\textbf{$CB_{20}$=\{ACC,ECG,EDA\}} & 69.46 & 74.81 &	84.11 & 85.62 & \cellcolor{gray!20}\textbf{84.0} & \cellcolor{gray!20}\textbf{86.37} & 66.96 & 73.53 & 73.19 &	76.18\\
\hline\textbf{$CB_{21}$=\{ECG,RESP,EMG\}} & 61.32 & 65.25 &	68.82 & 73.78 & 67.5 & 74.26 & 75.91 & 80.2 & 65.96 &	72.18\\
\hline\textbf{$CB_{22}$=\{ECG,EDA,TEMP\}} & 73.09 & 75.06 &	78.33 & 79.4 & 81.01 & 82.4 & 52.84 & 55.01 & 77.92 &	80.7\\
\hline\textbf{$CB_{23}$=\{ECG,RESP,EDA\}} & 70.15 & 72.45 &	78.83 & 80.05 & 79.93 & 81.62 & 67.9 & 74.41 & 76.45 &	79.31\\
\hline\textbf{$CB_{24}$=\{ECG,EMG,EDA\}} & 74.75 & 77.26 &	80.69 & 82.39 & 82.03 & 84.58 & 69.03 & 74.81 & 71.75 &	75.88\\
\hline\textbf{$CB_{25}$=\{RESP,EMG,EDA\}} & 60.6 & 65.44 &	74.27 & 78.99 & 73.3 & 79.62 & 54.6 & 62.95 & 67.22 &	76.13\\
\hline\textbf{$CB_{26}$=\{RESP,EDA,TEMP\}} & 69.55 & 71.23 &	77.57 & 78.93 & 77.44 & 79.83 & 35.86 & 39.07 & 70.25 &	75.78\\
\hline\textbf{$CB_{27}$=\{EMG,EDA,TEMP\}} & 70.92 & 73.2 &	80.65 & 83.39 & 77.88 & 82.14 & 50.88 & 54.17 & 66.11 &	73.9\\
\hline\textbf{$CB_{28}$=\{ACC,RESP\}} & 64.32 & 70.8 &	68.92 & 77.16 & 66.29 & 75.53 & 62.96 & 75.13 & 65.24 &	72.01\\
\hline\textbf{$CB_{29}$=\{ACC,EMG\}} & 61.76 & 69.37 &	66.36 & 76.85 & 64.19 & 75.86 & 60.91 & 69.19 & 60.51 &	68.4\\
\hline\textbf{$CB_{30}$=\{ACC,ECG\}} & 63.41 & 70.31 &	71.71 & 78.03 & 75.3 & 80.99 & 74.32 & 79.37 & 62.87 &	66.02\\
\hline\textbf{$CB_{31}$=\{ACC,EDA\}} & 65.63 & 73.35 &	77.51 & 81.63 & 69.97 & 78.57 & 50.03 & 59.14 & 64.28 &	70.83\\
\hline\textbf{$CB_{32}$=\{ACC,TEMP\}} & 65.69 & 72.52 &	76.55 & 82.0 & 68.5 & 78.26 & 73.3 & 76.77 & 67.49 &	74.41\\
\hline\textbf{$CB_{33}$=\{ECG,RESP\}} & 65.21 & 69.46 &	72.23 & 77.74 & 73.98 & 78.63 & 77.57 & 82.62 & 68.9 &	72.35\\
\hline\textbf{$CB_{34}$=\{ECG,EMG\}} & 59.41 & 64.16 &	68.24 & 73.16 & 66.69 & 74.49 & 75.07 & 79.67 & 65.89 &	71.05\\
\hline\textbf{$CB_{35}$=\{ECG,EDA\}} & 73.98 & 75.83 &	79.27 & 80.54 & 79.02 & 80.46 & 66.44 & 73.23 & 74.62 &	77.1\\
\hline\textbf{$CB_{36}$=\{ECG,TEMP\}} & 60.01 & 62.86 &	74.62 & 77.53 & 76.0 & 79.02 & 77.6 & 79.43 & 69.8 &	72.64\\
\hline\textbf{$CB_{37}$=\{EDA,TEMP\}} & 67.64 & 69.64 &	76.97 & 78.79 & 73.82 & 76.4 & 42.8 & 45.87 & 68.43 &	72.55\\
\hline\textbf{$CB_{38}$=\{RESP,EDA\}} & 56.46 & 59.02 &	74.86 & 77.79 & 66.92 & 71.09 & 48.47 & 59.08 & 71.71 &	78.38\\
\hline\textbf{$CB_{39}$=\{RESP,EMG\}} & 57.13 & 66.44 &	53.69 & 69.12 & 54.17 & 69.16 & 61.04 & 70.26 & 57.0 &	69.98\\
\hline\textbf{$CB_{40}$=\{RESP,TEMP\}} & 54.42 & 56.14 &	73.9 & 77.39 & 73.96 & 77.51 & 73.24 & 76.48 & 70.44 &	74.84\\
\hline\textbf{$CB_{41}$=\{EMG,EDA\}} & 61.9 & 67.19 &	76.18 & 80.4 & 72.54 & 79.26 & 51.34 & 59.27 & 66.06 &	74.2\\
\hline\textbf{$CB_{42}$=\{EMG,TEMP\}} & 56.15 & 60.49 &	77.06 & 81.6 & 72.14 & 78.9 & 73.38 & 76.76 & 62.5 &	71.6\\
\hline
\end{tabular}
\end{table*}

\begin{figure*} [!ht]
    \centering
    \includegraphics[width=16cm]{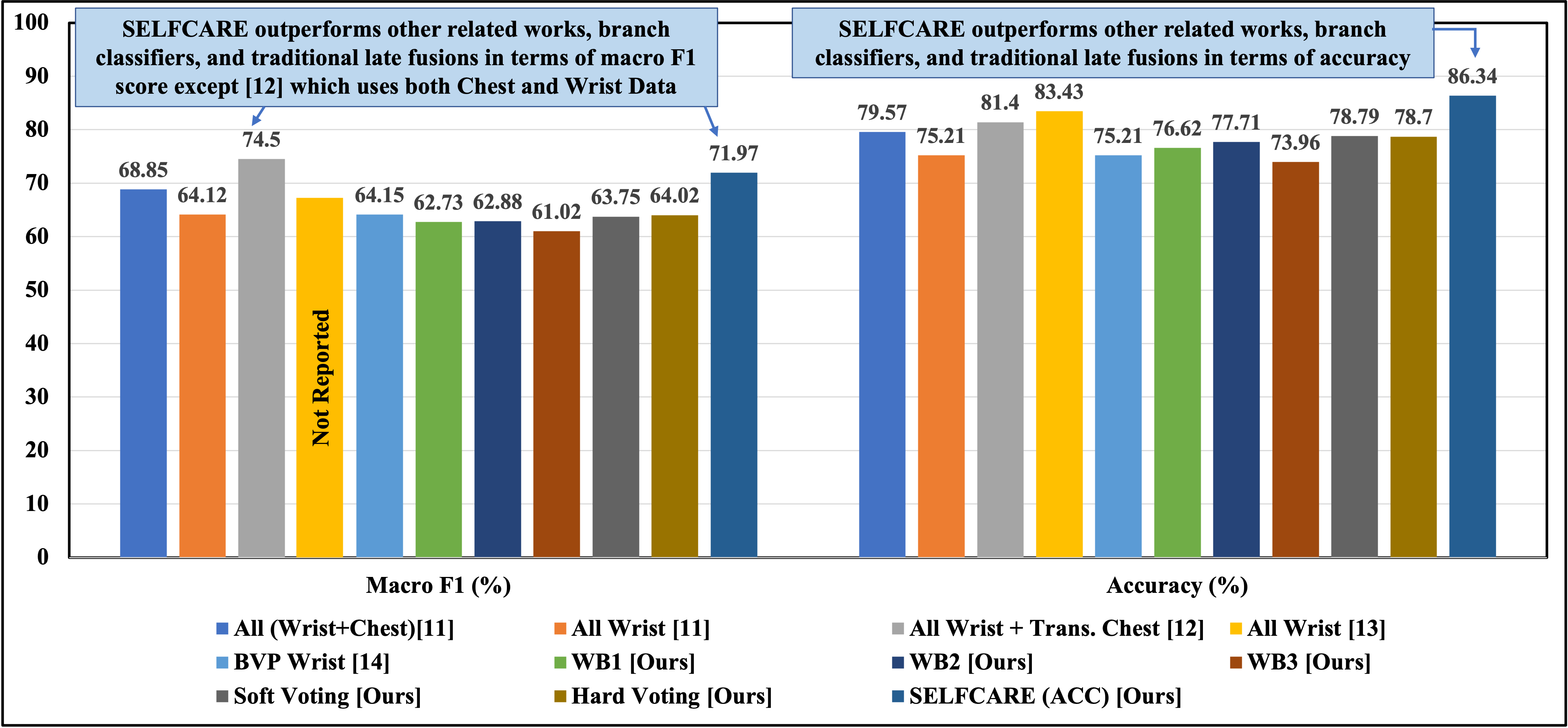}
    \caption{Overall performance comparison of related works using LOSO validation on wrist data 3-class. Results show that SELF-CARE outperforms the related works, branch classifiers, and other traditional late fusion methods in terms of macro F1 and accuracy except for the macro F1 score of \cite{samyoun2020stress} which uses both wrist and chest data.}
    \label{fig:overall_performance_wrist_3_class}
\end{figure*}

\subsubsection{Train Gating Model}
The gating model interprets the context of a sample by modeling the movement (for wrist-worn devices) or muscle contraction (for chest-worn devices) that occurred during that segment.
Therefore, we use the ACC (wrist) or EMG (chest) features as input data to train the gating model with the labels generated from the previous Section \ref{sec:generate_gating_label}.
We use a DT classifier as the gating model where the minimum number of samples to split a node is set to 20. The DT classifier is very lightweight and helps to minimize the overhead of SELF-CARE. Once the gating model is trained, the test subject data is used to test our architecture as shown in Fig. \ref{fig:arch}. For wrist-worn devices, the gating model outputs the probability of using the three final branch classifiers based on the test subject's ACC features. Similarly, for chest-worn devices, EMG features are  used by the gating model to determine the probability of using the five final branch classifiers as mentioned in Section \ref{sec:branch_classifier_selection}. One, two, or all of the final classifiers may be selected for final classification depending on the value of $\delta$, as discussed earlier in Section \ref{sec:delta}. For our 3-class (2-class) classification using wrist-worn devices, we set $\delta=0.40$ ($\delta=0.10$). And for the chest-worn devices, we set $\delta=0.20$ for 3-class and $\delta=0.15$ for 2-class classification. The model extracts additional features based on the required input of the selected branch classifiers, and applies a late fusion method to the classification output of the selected branches to generate the final result.

\subsubsection{Kalman Filter Tuning}
\label{sec:kf_tune}
The Kalman filter-based method is the only late fusion method in our implementation that requires tuning. As described in Section \ref{sec:lfb}, Kalman filters require an initial state ($\mathbf{x}_0$), state covariance ($\mathbf{P}_0$) and process noise and measurement noise vectors, $\mathbf{v}$ and $\mathbf{w}$, respectively. For the 3-class (2-class) classification using wrist-worn devices, we initialize $x_0 = [0.8,0.1,0.1]^\top$ ($x_0 = [0.8,0.2]^\top$). Similarly, for the 3-class and 2-class classification using chest-worn devices, $x_0$ is initialized to $[0.93,0.21,0.01]^\top$  $[1.0,0.55]^\top$. For 3-class (2-class) classification, we initialize $P_0 = 0.01\cdot\mathbf{I}_{3x3}$ ($P_0 = 0.01\cdot\mathbf{I}_{2x2}$) for both  wrist-worn and chest-worn devices. The state transition matrix $\mathbf{F}$ and measurement matrix $\mathbf{H}$ are identity matrices for the respective problems. The $\mathbf{Q}$ for both problems is modeled as a discrete time white process noise with variance set at 5e-4. The measurement noise is modeled as a function of each measurement to allow the filter to adjust the confidence of the measurements according to each reported class probability: $\mathbf{R} = ((\mathbf{1}-\mathbf{z})\cdot2\cdot\mathbf{I}_{3x3})^2$ ($\mathbf{R} = ((\mathbf{1}-\mathbf{z})/2\cdot\mathbf{I}_{2x2})^2$).
Lastly, a tunable threshold technique was used to process the measurements which involved (i) an $\epsilon$ parameter to select measurements which had a maximum predicted probability above the threshold and (ii) a $\gamma$ factor to scale the measurements to account for the imbalanced class distribution in the dataset. This thresholding process allows the filter to weigh each measurement it receives with a different degree of noise while also attempting to resolve issues that arise from imbalanced datasets. For the 3-class (2-class) classification using wrist-worn devices, we set $\epsilon = 0.4$ ($\epsilon = 0.7$) and $\gamma = [0.278,1,1]^\top $ ($\gamma = [0.667,1.1]^\top$). 
For the 3-class (2-class)  classification using chest-worn  devices, we set $\epsilon = 0.5$ ($\epsilon = 0.5$) and $\gamma = [1.35,1.5,1.6]^\top $ ($\gamma = [0.915,0.995]^\top$).


\newcommand{\bad}[0]{\cellcolor{red!10}}
\newcommand{\good}[0]{\cellcolor{green!10}}

\begin{figure*} [t]
    \centering
    \includegraphics[width=16cm]{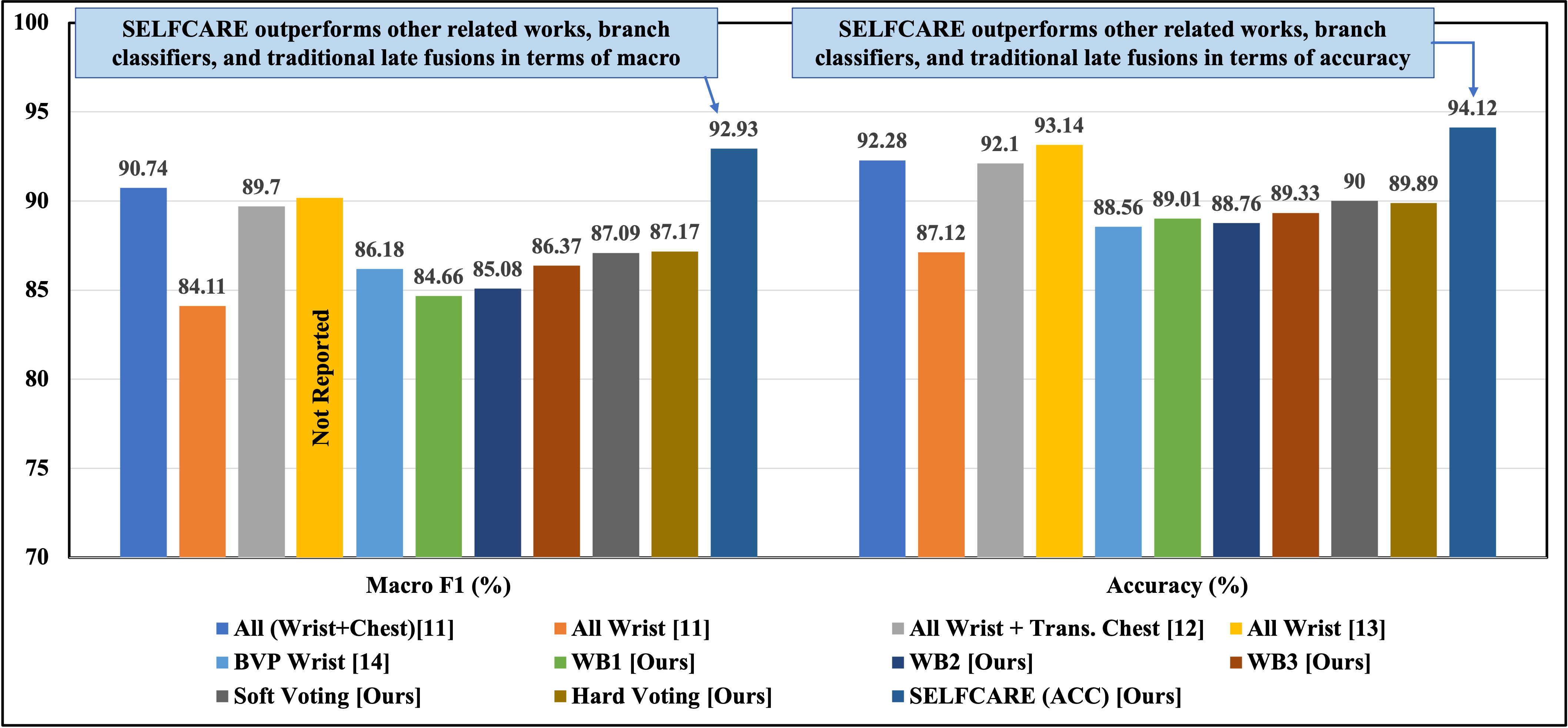}
    \caption{Overall performance comparison of related works using LOSO validation on wrist data 2-Class. Results show that SELF-CARE outperforms the related works, branch classifiers, and other traditional late fusion methods in terms of both macro F1 and accuracy.}
    \label{fig:overall_performance_wrist_2_class}
\end{figure*}
\subsection{Evaluation Metrics}
As stated previously, the WESAD dataset is highly imbalanced in terms of the number of segments per class. For this reason, we rely on both the F1 score and the accuracy to measure the classification performance. To ensure a fair comparison with other works, we use the macro F1 score. The metrics used for evaluation:
\begin{equation}
Accuracy = (TP+TN)/(TP+FP+TN+FN)
\end{equation}
\vspace{-5mm}
\begin{equation}
P = TP/(TP+FP), \; R = TP/(TP+FN)
\end{equation}
\vspace{-5mm}
\begin{equation}
Macro~F_1 = \frac {1} {{n_c}} \sum_{i}^{n_c} 2* \frac {P_i.R_i}{P_i+R_i}
\end{equation}
where TP, TN, FP, FN represents True Positives, True Negatives, False Positives, and False Negatives, respectively; and P and R represent Precision and Recall, respectively. The classes are indexed by \textit{i}, and $n_c$ is the number of output classes.

\begin{figure*} [t]
    \centering
    \includegraphics[width=16cm]{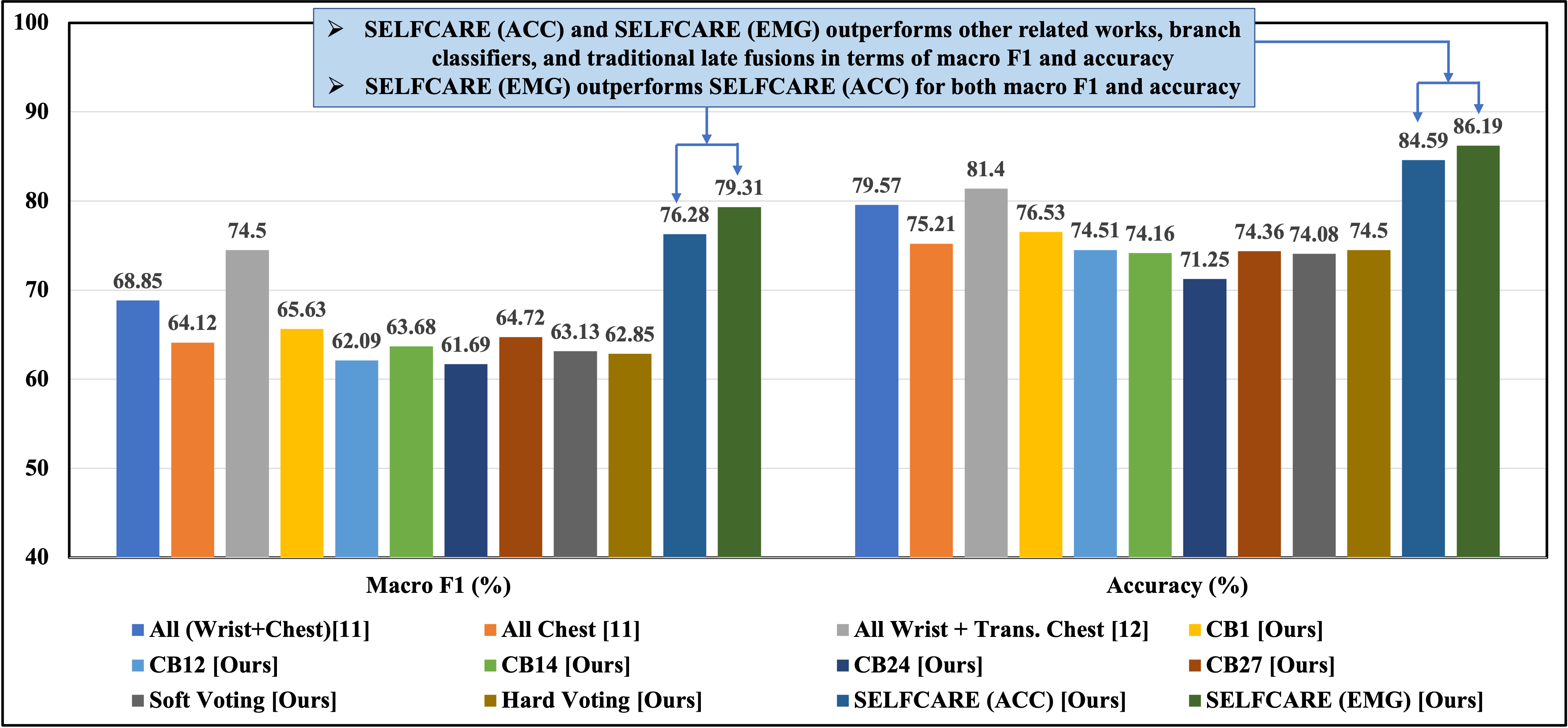}
    \caption{Overall performance comparison of related works using LOSO validation on chest data 3-class. Results show that the proposed SELF-CARE method (using either ACC or EMG to determine the noise context) outperforms other related works, branch classifiers, and traditional late fusion methods in terms of macro F1 and accuracy. Moreover, SELF-CARE using EMG (for the noise context understanding) outperforms the one using ACC in both metrics. This justifies the fact that EMG is more suitable than ACC to understand the noise context in chest wearable devices.}
    \label{fig:overall_performance_chest_3_class}
\end{figure*}

\begin{figure*} [!ht]
    \centering
    \includegraphics[width=16cm]{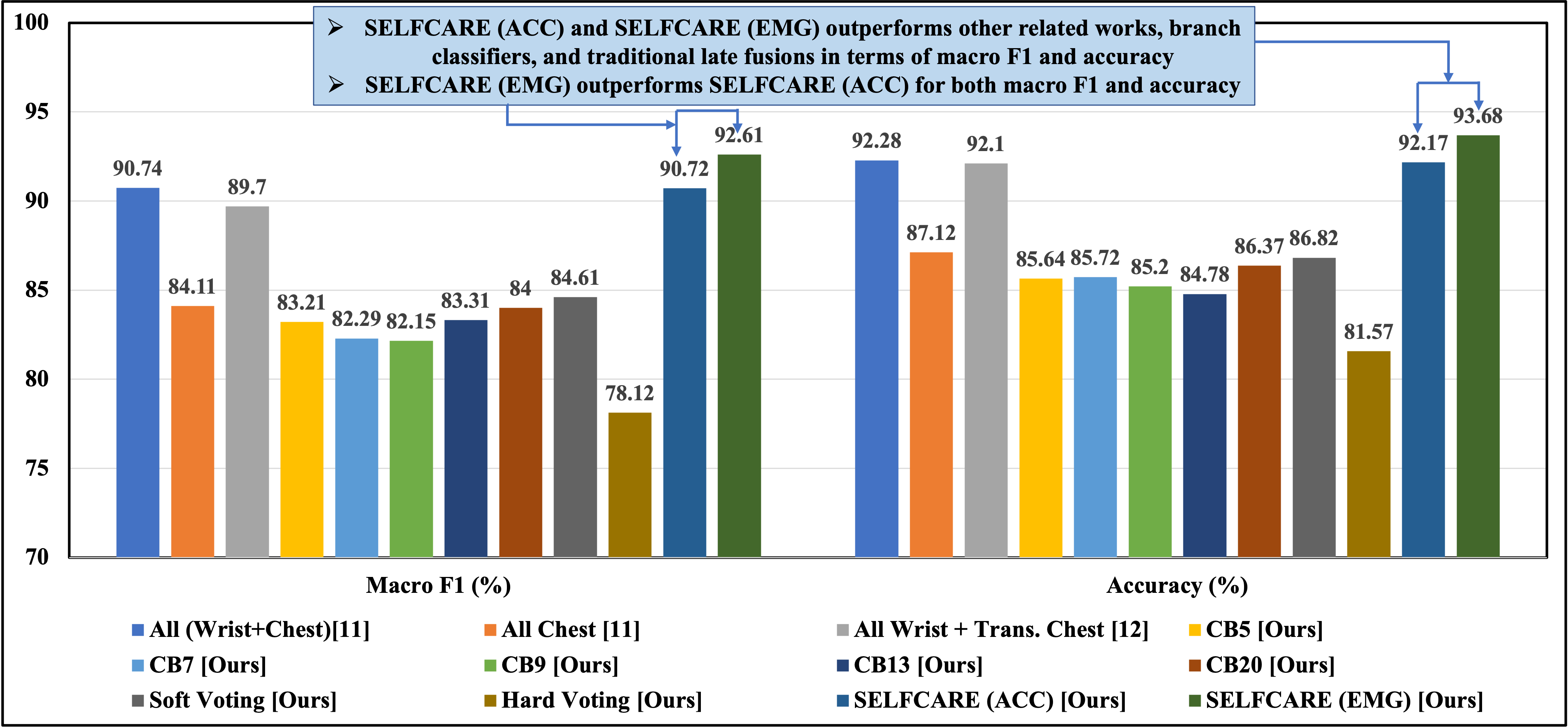}
    \caption{Overall performance comparison of related works using LOSO validation on chest data 2-class. Results show that the proposed SELF-CARE method outperforms other related works, branch classifiers, and traditional late fusion methods in terms of macro F1 and accuracy. Also, SELF-CARE using EMG outperforms the one using ACC in both metrics which further justifies the fact that EMG is more suitable than ACC to understand the noise context in chest wearable devices.}
    \label{fig:overall_performance_chest_2_class}
\end{figure*}

\subsection{Experimental Results}
This section presents the performance of SELF-CARE for stress detection in 3-class and 2-class classification using wrist and chest modalities. 


\subsubsection{Performance Evaluation of Wrist Modalities}
\label{wrist_performance_evaluation}

Tables \ref{tab:wrist_3_class_analysis} and \ref{tab:wrist_2_class_analysis} show the performance analysis of different classifiers for various input branches for the 3-class and 2-class problems, with each branch representing different combinations of input sensors. The RF classifier for branches $WB_1$, $WB_2$, and $WB_3$ shows better or competitive performance compared to the other classifiers for both 3-class and 2-class. The RF classifiers also achieved minimum training loss for these input branches during training, leading to our selection of these three branches with the RF classifier for our approach.

As shown in Fig. \ref{fig:overall_performance_wrist_3_class}, for 3-class classification, the SELF-CARE method outperforms other related works \cite{schmidt2018introducing,huynh2021stressnas,rashid2021feature}, the branch  classifiers, and the traditional late  fusion methods in terms of both accuracy and macro F1 score achieving a performance of 86.34\% and 71.97\%, respectively. Compared to \cite{samyoun2020stress}, SELF-CARE achieves better accuracy---though \cite{samyoun2020stress} achieves a better macro F1 score, as this work uses both wrist and chest sensors for stress classification. For 2-class classification, the SELF-CARE method achieves an accuracy of 94.12\% and macro F1 score of 92.93\%, outperforming  the related works \cite{schmidt2018introducing,huynh2021stressnas,rashid2021feature,samyoun2020stress}, the  branch  classifiers, and the traditional late fusion methods in terms of both accuracy and macro F1 score (as shown in Fig. \ref{fig:overall_performance_wrist_2_class}). For the three selected branch classifiers, we apply soft- and hard-voting methods, showing performance improvements compared to the individual branch classifiers for both 3-class and 2-class classifications. SELF-CARE also uses Kalman filter-based late fusion to further improve the performance for 3-class and 2-class classification compared to these traditional late fusion methods.

\subsubsection{Performance Evaluation of Chest Modalities}
\label{chest_performance_evaluation}

Tables \ref{tab:chest_3_class_analysis} and \ref{tab:chest_2_class_analysis} show the performance analysis of different classifiers for various input branches for the 3-class and 2-class problems, with each branch representing different combinations of input sensors. The AB classifier for branches $CB_1$, $CB_{12}$, $CB_{14}$, $CB_{24}$, and $CB_{27}$ shows better or competitive performance compared to the other classifiers 3-class classification. Similarly, for 2-class classification, the branches $CB_5$, $CB_7$, $CB_9$, $CB_{13}$, and $CB_{20}$ showed better performance that other classifiers. The AB classifiers also achieved minimum training loss for these input branches during training, which led to the selection of five branches for the SELF-CARE framework. The soft- and hard-voting methods applied to the five selected branch classifiers do not show performance improvements compared to the individual branch classifiers for both 3-class and 2-class classifications. However, incorporating Kalman filter-based late fusion significantly improves the performance for 3-class and 2-class classification compared to these traditional late fusion methods. 

As shown in Fig. \ref{fig:overall_performance_chest_3_class} and \ref{fig:overall_performance_chest_2_class}, for both 3-class and 2-class classification, the SELF-CARE method, using either muscle contraction (EMG) or motion (ACC) for context understanding, outperforms other related works \cite{schmidt2018introducing,huynh2021stressnas,rashid2021feature, samyoun2020stress}, the  branch  classifiers, and the traditional late  fusion methods in terms of both accuracy and macro F1 score. This study also demonstrates that even with motion-based context understanding, SELF-CARE outperforms other works. However, the model's performance improves by 2-3\% while  using  muscle contraction for context understanding compared to motion. This illustrates that the impact of movement on other sensors depends on the location of wearable devices. Therefore, movement is not always the best choice for contextual understanding as we observe the results while using chest modalities for stress detection.

\section{Limitations and Future Directions}
\label{sec:disc}
One of the main goals of this paper is to explore how the context of noise varies depending on the location of  wearable devices. 
For this reason, we modeled the noise context of sensor modalities from stand-alone devices, choosing not to combine the wrist and chest sensor modalities.
However, future research could explore this issue further.
Understanding the relation between the noise context of multiple wearable devices from physically different locations and fusing cross modal sensors based on that relation may produce interesting scientific findings that can be leveraged for methods of affective computing. 
Modeling the noise context of wearable health sensors can lead to further levels of human emotion understanding as information from the health sensors becomes increasingly useful when interpreted on a contextual basis.

Further, SELF-CARE is limited by the manual design of sensor fusion branch configurations. Though domain knowledge is required to determine which sensor data to fuse together, future works could explore using machine learning to make these determinations instead. Additionally, the energy efficiency of wearable health devices is an important constraint that could be examined in future works. SELF-CARE implements a configurable parameter for balancing computation and performance, but future works could examine the efficiency trade-offs between chest and wrist sensing modalities. This paper did not focus on deep learning models, but SELF-CARE's modular design allows for the implementation of any learning-based classifier, including deep learning branches. Further, SELF-CARE could be applied more broadly in the domain of affective computing to include additional tasks beyond stress detection and emotion recognition.
Moreover, it can also be applied to other wearable healthcare applications like human activity recognition \cite{rashid2022ahar, odema2021eexnas, rashid2021hear}, myocardial infarction detection \cite{rashid2020energy, odema2021energy, rashid2022template} etc., that involves data from multiple wearable sensors. Lastly, SELF-CARE's use of a specialized set of ensemble classifiers has broad applicability to IoT sensing, including the domains of sensor networks \cite{yasaei2020iot}, and transportation \cite{malawade2021sage}.


\section{Conclusion}
\label{sec:con}

In this paper, we propose SELF-CARE, a generalized selective sensor fusion method for stress detection that utilizes the noise context in the chest- and wrist-worn devices to dynamically adjust the sensor fusion performed to maximize classification performance. SELF-CARE determines the noise context using muscle contractions (EMG) or motion (ACC) of a subject, and performs an intelligent gating mechanism to select which sensor fusion schema to use depending on the location of the sensor.
 We also show that, while determining the noise context based on motion works best for wrist-based wearable devices, it is not the best for chest-based wearable devices. Through experimental evaluation, we conclude that EMG is better than ACC in understanding the noise context of chest-based wearable devices. To the best of our knowledge, SELF-CARE achieves state-of-the-art performance on the WESAD dataset for both chest and wrist-based sensors among methods that use LOSO validation. Using wrist-based sensors our methodology achieves 86.34\% (3-class) and 94.12\% (2-class) classification accuracy while outperforming current state-of-the-art works. Similarly, for chest-based wearable sensors, our methodology outperforms existing models with 86.19\% (3-class) and 93.68\% (2-class) classification accuracy. 


\balance
\bibliographystyle{IEEEtran}
\bibliography{bibliography}

\begin{thebibliography}{10}
\providecommand{\url}[1]{#1}
\csname url@samestyle\endcsname
\providecommand{\newblock}{\relax}
\providecommand{\bibinfo}[2]{#2}
\providecommand{\BIBentrySTDinterwordspacing}{\spaceskip=0pt\relax}
\providecommand{\BIBentryALTinterwordstretchfactor}{4}
\providecommand{\BIBentryALTinterwordspacing}{\spaceskip=\fontdimen2\font plus
\BIBentryALTinterwordstretchfactor\fontdimen3\font minus
  \fontdimen4\font\relax}
\providecommand{\BIBforeignlanguage}[2]{{%
\expandafter\ifx\csname l@#1\endcsname\relax
\typeout{** WARNING: IEEEtran.bst: No hyphenation pattern has been}%
\typeout{** loaded for the language `#1'. Using the pattern for}%
\typeout{** the default language instead.}%
\else
\language=\csname l@#1\endcsname
\fi
#2}}
\providecommand{\BIBdecl}{\relax}
\BIBdecl

\bibitem{american2020stress}
\BIBentryALTinterwordspacing
A.~P. Association. (2020) Stress in {A}merica 2020: a national mental health
  crisis. [Online]. Available:
  \url{https://www.apa.org/news/press/releases/stress/2020/sia-mental-health-crisis.pdf}
\BIBentrySTDinterwordspacing

\bibitem{goldstein2010adrenal}
D.~S. Goldstein, ``Adrenal responses to stress,'' \emph{Cellular and molecular
  neurobiology}, vol.~30, no.~8, pp. 1433--1440, 2010.

\bibitem{picard2000affective}
R.~W. Picard, \emph{Affective computing}.\hskip 1em plus 0.5em minus
  0.4em\relax MIT press, 2000.

\bibitem{affective_computing_market}
\BIBentryALTinterwordspacing
Markets and Markets. (2020) Affective computing market. [Online]. Available:
  \url{https://www.marketsandmarkets.com/Market-Reports/affective-computing-market-130730395.html}
\BIBentrySTDinterwordspacing

\bibitem{kim2008emotion}
J.~Kim and E.~Andr{\'e}, ``Emotion recognition based on physiological changes
  in music listening,'' \emph{IEEE transactions on pattern analysis and machine
  intelligence}, vol.~30, no.~12, pp. 2067--2083, 2008.

\bibitem{gjoreski2016continuous}
M.~Gjoreski, H.~Gjoreski, M.~Lu{\v{s}}trek, and M.~Gams, ``Continuous stress
  detection using a wrist device: in laboratory and real life,'' in
  \emph{proceedings of the 2016 ACM international joint conference on pervasive
  and ubiquitous computing: Adjunct}, 2016, pp. 1185--1193.

\bibitem{hovsepian2015cstress}
K.~Hovsepian, M.~Al'Absi, E.~Ertin, T.~Kamarck, M.~Nakajima, and S.~Kumar,
  ``cstress: towards a gold standard for continuous stress assessment in the
  mobile environment,'' in \emph{Proc. of the ACM {I}nt. joint conference on
  pervasive and ubiquitous computing}, 2015, pp. 493--504.

\bibitem{healey2005detecting}
J.~A. Healey and R.~W. Picard, ``Detecting stress during real-world driving
  tasks using physiological sensors,'' \emph{IEEE {T}rans. on intelligent
  transportation systems}, vol.~6, no.~2, pp. 156--166, 2005.

\bibitem{picard2001toward}
R.~W. Picard, E.~Vyzas, and J.~Healey, ``Toward machine emotional intelligence:
  Analysis of affective physiological state,'' \emph{IEEE {T}rans. on pattern
  analysis and machine intelligence}, vol.~23, no.~10, pp. 1175--1191, 2001.

\bibitem{koelstra2011deap}
S.~Koelstra, C.~Muhl, M.~Soleymani, J.-S. Lee, A.~Yazdani, T.~Ebrahimi, T.~Pun,
  A.~Nijholt, and I.~Patras, ``Deap: A database for emotion analysis; using
  physiological signals,'' \emph{IEEE transactions on affective computing},
  vol.~3, no.~1, pp. 18--31, 2011.

\bibitem{schmidt2018introducing}
P.~Schmidt, A.~Reiss, R.~Duerichen, C.~Marberger, and K.~Van~Laerhoven,
  ``Introducing wesad, a multimodal dataset for wearable stress and affect
  detection,'' in \emph{Proceedings of the 20th ACM international conference on
  multimodal interaction}, 2018, pp. 400--408.

\bibitem{samyoun2020stress}
S.~Samyoun, A.~S. Mondol, and J.~A. Stankovic, ``Stress detection via sensor
  translation,'' in \emph{2020 16th International Conference on Distributed
  Computing in Sensor Systems (DCOSS)}.\hskip 1em plus 0.5em minus 0.4em\relax
  IEEE, 2020, pp. 19--26.

\bibitem{huynh2021stressnas}
L.~Huynh, T.~Nguyen, T.~Nguyen, S.~Pirttikangas, and P.~Siirtola, ``Stressnas:
  Affect state and stress detection using neural architecture search,'' in
  \emph{Adj. Proc. of the 2021 ACM International Joint Conference on Pervasive
  and Ubiquitous Computing and Proc. of the 2021 ACM International Symposium on
  Wearable Computers}, 2021, pp. 121--125.

\bibitem{rashid2021feature}
N.~Rashid, L.~Chen, M.~Dautta, A.~Jimenez, P.~Tseng, and M.~A. Al~Faruque,
  ``Feature augmented hybrid cnn for stress recognition using wrist-based
  photoplethysmography sensor,'' in \emph{2021 43rd Annual International
  Conference of the IEEE Engineering in Medicine \& Biology Society
  (EMBC)}.\hskip 1em plus 0.5em minus 0.4em\relax IEEE, 2021, pp. 2374--2377.

\bibitem{ragav2019scalable}
A.~Ragav, N.~H. Krishna, N.~Narayanan, K.~Thelly, and V.~Vijayaraghavan,
  ``Scalable deep learning for stress and affect detection on
  resource-constrained devices,'' in \emph{2019 18th IEEE Int. Conference On
  Machine Learning And Applications (ICMLA)}.\hskip 1em plus 0.5em minus
  0.4em\relax IEEE, 2019, pp. 1585--1592.

\bibitem{schmidt2019multi}
P.~Schmidt, R.~D{\"u}richen, A.~Reiss, K.~Van~Laerhoven, and T.~Pl{\"o}tz,
  ``Multi-target affect detection in the wild: an exploratory study,'' in
  \emph{Proc. of the 23rd {I}nt Symposium on Wearable Computers}, 2019, pp.
  211--219.

\bibitem{naeini2021amser}
E.~K. Naeini, S.~Shahhosseini, A.~Kanduri, P.~Liljeberg, A.~M. Rahmani, and
  N.~Dutt, ``Amser: Adaptive multi-modal sensing for energy efficient and
  resilient ehealth systems,'' \emph{arXiv preprint arXiv:2112.08176}, 2021.

\bibitem{bota2020emotion}
P.~Bota, C.~Wang, A.~Fred, and H.~Silva, ``Emotion assessment using feature
  fusion and decision fusion classification based on physiological data: Are we
  there yet?'' \emph{Sensors}, vol.~20, no.~17, p. 4723, 2020.

\bibitem{malawade2022hydrafusion}
A.~V. Malawade, T.~Mortlock, and M.~A.~A. Faruque, ``Hydrafusion: Context-aware
  selective sensor fusion for robust and efficient autonomous vehicle
  perception,'' \emph{arXiv preprint arXiv:2201.06644}, 2022.

\bibitem{rashid2022self}
N.~Rashid, T.~Mortlock, and M.~A. Al~Faruque, ``Self-care: Selective fusion
  with context-aware low-power edge computing for stress detection,'' in
  \emph{2022 18th International Conference on Distributed Computing in Sensor
  Systems (DCOSS)}.\hskip 1em plus 0.5em minus 0.4em\relax IEEE, 2022, pp.
  49--52.

\bibitem{lin2019explainable}
J.~Lin, S.~Pan, C.~S. Lee, and S.~Oviatt, ``An explainable deep fusion network
  for affect recognition using physiological signals,'' in \emph{Proceedings of
  the 28th ACM International Conference on Information and Knowledge
  Management}, 2019, pp. 2069--2072.

\bibitem{fouladgar2021cn}
N.~Fouladgar, M.~Alirezaie, and K.~Fr{\"a}mling, ``Cn-waterfall: a deep
  convolutional neural network for multimodal physiological affect detection,''
  \emph{Neural Computing and Applications}, pp. 1--20, 2021.

\bibitem{gravina2017multi}
R.~Gravina, P.~Alinia, H.~Ghasemzadeh, and G.~Fortino, ``Multi-sensor fusion in
  body sensor networks: State-of-the-art and research challenges,''
  \emph{Information Fusion}, vol.~35, pp. 68--80, 2017.

\bibitem{oviatt2018handbook}
S.~Oviatt, B.~Schuller, P.~Cohen, D.~Sonntag, G.~Potamianos, and A.~Kr{\"u}ger,
  \emph{The handbook of multimodal-multisensor interfaces, Volume 2: Signal
  processing, architectures, and detection of emotion and cognition}.\hskip 1em
  plus 0.5em minus 0.4em\relax Morgan \& Claypool, 2018.

\bibitem{yan2022emotion}
M.~Yan, Z.~Deng, B.~He, C.~Zou, J.~Wu, and Z.~Zhu, ``Emotion classification
  with multichannel physiological signals using hybrid feature and adaptive
  decision fusion,'' \emph{Biomedical Signal Processing and Control}, vol.~71,
  p. 103235, 2022.

\bibitem{kalman1960new}
R.~E. Kalman, ``A new approach to linear filtering and prediction problems,''
  1960.

\bibitem{pakrashi2019kalman}
A.~Pakrashi and B.~Mac~Namee, ``Kalman filter-based heuristic ensemble (kfhe):
  A new perspective on multi-class ensemble classification using kalman
  filters,'' \emph{Information Sciences}, vol. 485, pp. 456--485, 2019.

\bibitem{salehizadeh2016novel}
S.~Salehizadeh, D.~Dao, J.~Bolkhovsky, C.~Cho, Y.~Mendelson, and K.~H. Chon,
  ``A novel time-varying spectral filtering algorithm for reconstruction of
  motion artifact corrupted heart rate signals during intense physical
  activities using a wearable photoplethysmogram sensor,'' \emph{Sensors},
  vol.~16, no.~1, p.~10, 2016.

\bibitem{rashid2022ahar}
N.~Rashid, B.~U. Demirel, and M.~A. Al~Faruque, ``Ahar: Adaptive cnn for
  energy-efficient human activity recognition in low-power edge devices,''
  \emph{IEEE Internet of Things Journal}, 2022.

\bibitem{odema2021eexnas}
M.~Odema, N.~Rashid, and M.~A. Al~Faruque, ``Eexnas: Early-exit neural
  architecture search solutions for low-power wearable devices,'' in \emph{2021
  IEEE/ACM International Symposium on Low Power Electronics and Design
  (ISLPED)}.\hskip 1em plus 0.5em minus 0.4em\relax IEEE, 2021, pp. 1--6.

\bibitem{rashid2021hear}
N.~Rashid, M.~Dautta, P.~Tseng, and M.~A. Al~Faruque, ``Hear: Fog-enabled
  energy-aware online human eating activity recognition,'' \emph{IEEE Internet
  of Things Journal}, vol.~8, no.~2, pp. 860--868, 2021.

\bibitem{rashid2020energy}
N.~Rashid and M.~A. Al~Faruque, ``Energy-efficient real-time myocardial
  infarction detection on wearable devices,'' in \emph{2020 42nd Annual
  International Conference of the IEEE Engineering in Medicine \& Biology
  Society (EMBC)}.\hskip 1em plus 0.5em minus 0.4em\relax IEEE, 2020, pp.
  4648--4651.

\bibitem{odema2021energy}
M.~Odema, N.~Rashid, and M.~A. Al~Faruque, ``Energy-aware design methodology
  for myocardial infarction detection on low-power wearable devices,'' in
  \emph{Proceedings of the 26th Asia and South Pacific Design Automation
  Conference}, 2021, pp. 621--626.

\bibitem{rashid2022template}
\BIBentryALTinterwordspacing
N.~Rashid, B.~U. Demirel, M.~Odema, and M.~A. Al~Faruque, ``Template matching
  based early exit cnn for energy-efficient myocardial infarction detection on
  low-power wearable devices,'' \emph{Proc. ACM Interact. Mob. Wearable
  Ubiquitous Technol.}, vol.~6, no.~2, jul 2022. [Online]. Available:
  \url{https://doi.org/10.1145/3534580}
\BIBentrySTDinterwordspacing

\bibitem{yasaei2020iot}
R.~Yasaei, F.~Hernandez, and M.~A.~A. Faruque, ``Iot-cad: Context-aware
  adaptive anomaly detection in iot systems through sensor association,'' in
  \emph{Proceedings of the 39th International Conference on Computer-Aided
  Design}, 2020, pp. 1--9.

\bibitem{malawade2021sage}
A.~Malawade, M.~Odema, S.~Lajeunesse-DeGroot, and M.~A. Al~Faruque, ``Sage: A
  split-architecture methodology for efficient end-to-end autonomous vehicle
  control,'' \emph{ACM Transactions on Embedded Computing Systems (TECS)},
  vol.~20, no.~5s, pp. 1--22, 2021.

\end{thebibliography}

\end{document}